\RequirePackage[l2tabu, orthodox]{nag}
\pdfoutput=1
\documentclass[a4paper,9pt,twocolumn]{article}

\usepackage[utf8]{inputenc}
\usepackage[T1]{fontenc}
\usepackage[english]{babel}
\usepackage{amsmath}
\usepackage{mathtools}
\usepackage{comment}
\usepackage{textcomp}
\usepackage[style=base]{caption}
\usepackage{bm}
\usepackage{url}
\usepackage{pdflscape}
\usepackage[margin=15mm,tmargin=2cm,bmargin=2.5cm]{geometry}
\usepackage{color}

\usepackage{setspace}

\usepackage[bitstream-charter]{mathdesign}

\usepackage{sectsty}
\sectionfont{\fontsize{12}{15}\selectfont}
\subsectionfont{\fontsize{10}{13}\selectfont}

\usepackage{csquotes}
\usepackage[
    backend=biber,
    citestyle=numeric,
    bibstyle=phys,
    articletitle=false,
    biblabel=brackets,
    chaptertitle=false,
    pageranges=false,
    isbn=false,
    url=false,
    sorting=none
]{biblatex}
\usepackage[symbolpackage=l3draw, symbol=plos]{biblatex-ext-oa}
\DeclareFieldFormat{titlecase}{#1}

\usepackage{overpic}
\usepackage[normalem]{ulem}

\newcommand{\mytitle}{Droplets on substrates with oscillating wettability}
\newcommand{\authors}{Josua Grawitter and Holger Stark}

\renewcommand{\vec}[1]{\bm{#1}}
\newcommand{\tens}[1]{\mathsf{#1}}
\newcommand{\dif}{\,\mathrm{d}}

\usepackage[
    colorlinks=false,
    pdfborder={0 0 0},
    linktoc=all,
    pdfpagelabels,
    plainpages=false,
    bookmarks=true,
    hypertexnames=false, 
    pdftex,
    pdfauthor={\authors},
    pdftitle={\mytitle}
]{hyperref}
\usepackage{cleveref}

\newcommand*{\halurl}[1]{http://hal.archives-ouvertes.fr/#1}
\DeclareFieldFormat{eprint:hal}{%
  \ifhyperref
    {\href{\halurl{#1}}{hal:~\nolinkurl{#1}}}
    {hal:~\nolinkurl{#1}}}
\DeclareFieldAlias{eprint:HAL}{eprint:hal}

\DeclareOpenAccessEprintUrl[always]{hal}{%
  http://hal.archives-ouvertes.fr/\thefield{eprint}}
\DeclareOpenAccessEprintAlias{HAL}{hal}

\newcommand*{\semschurl}[1]{https://api.semanticscholar.org/CorpusID:#1}
\DeclareFieldFormat{eprint:semanticscholar}{%
  \ifhyperref
    {\href{\semschurl{#1}}{hal:~\nolinkurl{#1}}}
    {hal:~\nolinkurl{#1}}}
\DeclareFieldAlias{eprint:SEMANTICSCHOLAR}{eprint:semanticscholar}

\DeclareOpenAccessEprintUrl[always]{semanticscholar}{%
  https://api.semanticscholar.org/CorpusID:\thefield{eprint}}
\DeclareOpenAccessEprintAlias{SEMANTICSCHOLAR}{semanticscholar}
\bibliography{rsc}

\usepackage{microtype}

\title{\mytitle}
\author{\authors}

\begin{document}
\renewcommand{\thefootnote}{\alph{footnote}}
\newgeometry{left=15mm,right=45mm}
\onecolumn
\noindent
{\huge \textbf{\mytitle}$^\dag$}

{\large

\vspace{0.5cm}
\noindent \textit{
Josua Grawitter\footnotemark[1]\textsuperscript{,}\footnotemark[2] and Holger Stark\footnotemark[1]\textsuperscript{,}\footnotemark[3]
}
\hfill 
July 30, 2021

\footnotetext[1]{Technische Universität Berlin, Institut für Theoretische Physik, Straße des 17.~Juni 135, 10623 Berlin, Germany.}
\footnotetext[2]{E-mail:~\href{mailto:josua.grawitter@physik.tu-berlin.de}{\nolinkurl{josua.grawitter@physik.tu-berlin.de}}}
\footnotetext[3]{E-mail:~\href{mailto:holger.stark@tu-berlin.de}{\nolinkurl{holger.stark@tu-berlin.de}}}
\footnotetext{\hspace{-4pt}$^\dag$Electronic Supplementary Information (ESI) available: Their descriptions are collected in Appendix~\ref{appendix_esi}.}

\renewcommand{\thefootnote}{\arabic{footnote}}

\vspace{1cm}

\begin{spacing}{1.15}
\noindent\textbf{Abstract}\newline
In recent decades novel solid substrates have been designed which change their wettability in response to light or an electrostatic field.
Here, we investigate a droplet on substrates with oscillating uniform wettability by varying minimium and maximum contact angles and frequency.
To simulate this situation, we use our previous work  [Grawitter and Stark, Soft Matter~\textbf{17}, 2454 (2021)], where we implemented the  boundary element method in combination with the Cox-Voinov law for the contact-line velocity, to determine the fluid flow inside a droplet.
After a transient regime the droplet performs steady oscillations, the amplitude of which decreases with increasing frequency.
For slow oscillations our numerical results agree well with the linearized spherical-cap model.
They collapse on a master curve when we rescale frequency by a characteristic relaxation time.
In contrast, for fast oscillations we observe significant deviations from the master curve.
The decay of the susceptibility is weaker and the phase shift between oscillations in wettability and contact angle stays below the predicted $\pi/2$.
The reason becomes obvious when studying the combined dynamics of droplet height and contact angle.
It reveals non-reciprocal shape changes during one oscillation period even at low frequencies due to the induced fluid flow inside the droplet, which are not captured by the spherical-cap model.
Similar periodic non-reciprocal shape changes occur at low frequencies when the droplet is placed on an oscillating nonuniform wettability profile with six-fold symmetry. 
Such profiles are inspired by the light intensity pattern of Laguerre-Gauss laser modes.
Since the non-reciprocal shape changes induce fluid circulation, which is controllable from the outside, our findings envisage to design targeted  microfluidic transport of solutes inside the droplet.
\end{spacing}
} 


\restoregeometry
\twocolumn
\section{Introduction}

The shape a liquid droplet forms on a flat surface is determined by the wettability landscape of the surface.
On more wettable parts of the surface, the droplet spreads out and on less wettable parts it contracts~\cite{bonn_wetting_2009}.
In the past two decades researchers have developed substrates the wettability of which can be controlled such that patterns that change in space and/or time can be created~\cite{ichimura_light_2000,seki_wide_2018,malinowski_advances_2020}.
For example, the wetted substrates become switchable due to a single layer of light-responsive molecules~\cite{ichimura_light_2000,delorme_azobenzene_2005,bunker_reversible_2008}, electro-responsive molecules~\cite{lahann_reversibly_2003}, an array of light-responsive pillars~\cite{jiang_photoswitched_2005,pirani_light_2016,oscurato_light_2017}, or more complex nano\-structures~\cite{lim_photoreversibly_2006,lim_uvdriven_2007}.
Regardless of the specific mechanism,  controlling the dynamics of a liquid drop  by time-varying wettability patterns is not yet fully explored, despite its importance for lab-on-a-chip devices~\cite{samiei_review_2016} and for targeted deposition of solutes, for exampe, in printing devices~\cite{mampallil_review_2018}.

So far, much research has aimed at understanding wetting on substrates with static non-uniform wettability patterns.
Early theoretical studies used perturbation methods to analytically estimate the influence of small local wettability gradients on droplets with simple circular or cylindrical shapes~\cite{greenspan_motion_1978,brochard_motions_1989}.
Later, experimental and numerical work investigated more complex shapes which occur, for example, when a droplet crosses a static step in wettability~\cite{glasner_boundary_2005,moosavi_size_2008}, flows over two neighboring stripes of increased wettability~\cite{dietrich_wetting_2005}, over a checker-board pattern~\cite{kaspar_confinement_2016,savva_droplet_2019}, or random spatial fluctuations in wettability~\cite{savva_droplet_2019}.
From the perspective of the droplet these patterns also become time-varying if the droplets starts to move, for example, on an inclined substrate~\cite{engelnkemper_continuation_2019}.
Furthermore, a droplet may trigger a chemical reaction with the substrate and thereby create a wettability gradient, which moves it forward~\cite{thiele_dynamical_2004}.
Taking into account the switchable substrates introduced above, we have shown recently how a chemically inert droplet responds to \emph{moving} steps in wettability~\cite{grawitter_steering_2021}.
Thus, the droplet's motion is under \emph{external} control.

In this article we present a theoretical investigation how a small liquid droplet behaves on a substrate with oscillations in \emph{uniform} wettability.
We also give a brief outlook toward its behavior on a substrate with oscillations of a \emph{non-uniform} wettability pattern.
Generally, small sessile droplets form spherical caps on flat substrates with uniform wettability because that shape minimizes its total surface energy on a flat substrate~\cite{degennes_wetting_1985}.
However, when the substrate's wettability oscillates, the droplet continually tries to follow but cannot relax to its equilibrium shape.
Its continuous motion in turn gives rise to internal flow in the droplet.
We are interested in both---the droplet shape during oscillation and the accompanying internal flow.
Since at small length scales viscous forces dominate inertia within the droplet, the internal fluid flow is described by the Stokes equations~\cite{huh_hydrodynamic_1971}.
We solve these using the boundary element method, the implementation of which we have described in a previous work~\cite{grawitter_steering_2021}.
At the edge of the droplet-substrate interface (the \emph{contact line}), we use the Cox-Voinov law to calculate the velocity of the contact line~\cite{voinov_hydrodynamics_1976}.
We have previously validated and applied our method to droplets that are
steered by moving steps in wettability~\cite{grawitter_steering_2021}.

The case of droplets exposed to oscillating wettability is distinct from droplets on vertically vibrated substrates, which have been studied in Refs.~\cite{noblin_vibrations_2009,sudo_dynamic_2010,singla_dynamics_2019}.
For the droplet such vibrations only play a role in the presence of inertia or external forces.
They affect the droplet as a whole, while oscillations in wettability only act via forces at the contact line.
Vertical vibrations have been shown to give rise to ripples that travel up the side of the droplet and to generate higher-harmonic deformations of the contact line for water~\cite{noblin_vibrations_2009,sudo_dynamic_2010} as well as for mercury droplets~\cite{singla_dynamics_2019}.
At large amplitudes of the vibrations the droplet breaks up by ejecting
small amounts of liquid at its top~\cite{sudo_dynamic_2010}.
As we study wettability oscillations in the absence of inertia, we will not observe such extreme phenomena in our case.

Our theoretical approach stands alongside two other continuum approaches to dynamic wetting.
In the first approach, one uses a thin-film equation to evaluate the droplet dynamics via its height profile~\cite{teletzke_wetting_1988,oron_long_1997,popescu_precursor_2012,thiele_recent_2018}, which means the contact angle should be small and cannot exceed 90~degrees.
Another approach, which we will discuss in detail below, is the \emph{spherical cap model}.
It constraints the shape of the droplet  to a spherical cap~\cite{voinov_hydrodynamics_1976,deruijter_contact_1997} and does not capture fluid flow within the droplet.
The spherical cap shape is motivated by the equilibrium shape of a droplet on a substrate with uniform wettability.
With our approach, we are able to evaluate the applicability of the spherical cap model and investigate the internal flow field of the droplet.

Our findings add to the microfluidics toolbox~\cite{squires_microfluidics_2005,baigl_photo_2012,dai_directional_2021} another way to interact with and manipulate droplets by placing them on substrates with oscillating wettability.
Specifically, we find that the contact angle oscillations of the droplet decrease with increasing frequency.
For slow oscillations this can be well described by the spherical-cap model, which even provides a characteristic time scale to map the oscillations onto a common master curve.
However, the master curve is no longer applicable for fast oscillations.
A more detailed study of the droplet dynamics in terms of two shape variables, such as contact angle and droplet height, reveals that they oscillate out-of-phase with each other.
Thus, the droplet performs a non-reciprocal motion during one oscillation period, which cannot be described by the spherical-cap model.
It is due to fluid flow within the droplet, which gives rise to fluid circulation  within the droplet.

Our article is structured as follows:
In Sect.~\ref{sec_method} we reiterate the theoretical basis of our boundary element method applied to dynamic wetting.
In Sect.~\ref{sec_main} we describe and discuss our findings in detail.
First, in Sect.~\ref{sec_phenom} we present the basic phenomenology of the droplet oscillations.
We analyze them using linear-response theory and the spherical-cap model, where the contact angle serves as a single shape characteristic.
Second, in Sect.~\ref{sec_pumping} we look at the coupled dynamics of contact angle and height, reveal the non-reciprocal motion of the droplet, and discuss its implications for the internal fluid flow.
Third, in Sect.~\ref{sec_nonuniform} we study a closely-related example of a droplet on a substrate with an oscillating non-uniform wettability pattern.
Finally, we conclude in Sect.~\ref{sec_conclusions}.

\section{Simulation method}
\label{sec_method}
The motion of a droplet consisting of an incompressible simple liquid is completely described by the dynamics of its interfaces: the gas-liquid and the solid-liquid interface.
The motion of any point~$\vec s$ on the interfaces is governed by the fluid velocity field at this point:
\begin{equation}
 \vec{\dot{s}} = \vec v(\vec s)~.
\end{equation}
In the following we summarize how we determine the velocity field both at the droplet surface and its interior.

\subsection{Stokes flow}
We consider droplets in which viscous drag dominates inertia, which is the limit of Stokes or creeping flow \cite{kim_microhydrodynamics_2005}.

The equations govering the velocity field~$\vec v$ of Stokes flow are
\begin{align}
 \mu\nabla^2\vec v &= \nabla p, & \nabla\cdot \vec v &= 0~,
 \label{eq_stokes}
\end{align}
where $\mu$ is viscosity and the second equation is the incompressibility condition which constrains the pressure~$p$.
These differential equations can be restated as boundary integral equations~\cite{pozrikidis_boundary_1992} using the Oseen tensor~$\tens{O}$ and the associated stress field $\mathsf{T}$:
\begin{multline}
c(\vec r) \vec{v}(\vec{r}) =
    \oint\limits_{\partial D} \tens{O}(\vec{r-r'}) \vec{\sigma n}(r') \dif^2\vec{r'}\\
    -\oint\limits_{\partial D} \vec{v}(\vec{r'}) \cdot \tens{T}(\vec{r-r'}) \vec{n}(\vec{r'}) \dif^2\vec{r'}
    \label{eq_boundary}
\end{multline}
with
\begin{equation}
   c(\vec r) =
    \begin{cases}
    1 & \text{for } \vec r \in D \setminus \partial D
    \\
    \frac{1}{2} & \text{for } \vec r \in \partial D \text{, where $\partial D$ is smooth}
    \\
    \frac{\alpha}{4\pi} & \text{for } \vec r \in \partial D \text{, where $\partial D$ has a corner}
    \\
    &\text{with inward \textit{solid} angle $\alpha$.}
    \end{cases}
\end{equation}
where $\partial D$ is the (time-dependent) surface of the droplet.
Because the equation relates velocity $\vec v$ and stress $\vec{\sigma} \vec n$,  for any surface point $\vec r$ either variable must be prescribed by boundary conditions.

\subsection{Boundary conditions}

On the liquid-gas interface the normal stress balances surface tension forces due to mean curvature~$\kappa$ of the interface, \textit{i.e.}, $\vec {\sigma n} = \gamma \kappa \vec n$, where $\gamma$ is the surface tension of the liquid-gas interface.

On the solid liquid interface (at the substrate) two conditions apply:
Firstly, the interface cannot deform along its normal~$\vec e_z$ and therefore $v_z = 0$.
Secondly, roughness of the substrate introduces a small amount of slip with slip length~$\lambda$, which we account for by setting $\lambda \vec{\sigma n}= \mu\vec v$ tangential to the interface~\cite{bolanos_derivation_2017}.

As a boundary condition on the contact line, we choose its velocity along the substrate according to the Cox-Voinov law~\cite{voinov_hydrodynamics_1976,cox_dynamics_1986}
\begin{equation}
 v_\mathrm{contact} = \frac{\gamma}{9\mu \ln(h/\lambda)} (\theta_\mathrm{dyn}^3 - \theta_\mathrm{eq}^3) \, ,
 \label{eq_coxvoinov}
\end{equation}
where $\theta_\mathrm{eq}$ is the equilibrium contact angle, which defines the wettability of the substrate, and $\theta_\mathrm{dyn}$ is the dynamic or actual contact angle.
Only with this separate boundary condition is the problem well-posed because the three-phase contact line is neither clearly part of the gas-liquid interface nor the solid-liquid interface~\cite{moffatt_viscous_1964,huh_hydrodynamic_1971}.
Note also that the Cox-Voinov law excludes the effects of contact angle hysteresis~\cite{eral_contact_2013,snoeijer_moving_2013}.

The boundary conditions introduce several material parameters in addition to  viscosity $\mu$.
For our simulations we choose dimensionless parameters.
They correspond, for example, to a droplet with an initial radius~$R_0=100\,\mu\mathrm{m}$ of its circular base area and made of a 90\%~glycerol and 10\%~water mixture.
This reference system was studied in experiments by de Ruijter~\emph{et al.}~\cite{deruijter_contact_1997}.
The mixture has
$\mu=209\,\mathrm{mPa}\!\cdot\!\mathrm{s}$,
kinematic viscosity~$\nu= \mu/ \rho = 169\,\mathrm{mm}^2 \!\cdot\!\mathrm{s}^{-1}$,
$\gamma=65.3\,\mathrm{mN}\!\cdot\!\mathrm{m}^{-1}$,
$\lambda=1\,\mathrm{nm}$,
and $\ln(h/\lambda)=44$.
The latter value was observed by de~Ruijter \emph{et al.} by fitting the spherical cap model with Eqs.~(\ref{eq_spherical1}) and (\ref{eq_spherical2}) mentioned below to their experimental data.
Furthermore, we choose the initial contact angle~$\theta_\mathrm{dyn}$ to be the time average of $\theta_\mathrm{eq}$.
We calculate and report our data in units of $R_0$ for length, $\tau=R_0^2/\nu$ for time, and $F_0=\nu\mu$ for force.
Thus the remaining dimensionless parameters are $\tilde \gamma=\gamma R_0/F_0= 0.19$ and $\tilde \lambda = \lambda / R_0=10^{-5}$.

\subsection{Boundary element method}
To solve for the velocity field at the droplet's interfaces, we construct a triangular mesh and discretize the integral equation.
We then integrate the dynamic of the mesh in time using an adaptive 5th order Runge-Kutta method~\cite{tsitouras_runge_2011}.
The full details of our boundary element method are provided in Ref.~\cite{grawitter_steering_2021}. Briefly, to discretize 
Eq.~(\ref{eq_boundary}) we divide the droplet surface into polygonal regions, each with a vertex at its center. The polygons are then decomposed into triangles and we integrate separately over each triangle using Gaussian quadrature~\cite{katsikadelis_boundary_2016} 
with 400~sampling points for singular integrands and 9~sampling points for nonsingular integrands. Once we have solved the discretized equation and the surface velocity field is known, we can use the boundary integral equation~(\ref{eq_boundary}) to evaluate the flow field in the interior of the droplet. Similar numerical approaches have been used to study dewetting of polymer microdroplets~\cite{mcgraw_slip_2016,chan_morphological_2017} and for bubbles on a solid surface under the influence of an acoustic field~\cite{pityuk_boundary_2018}.

\section{Droplet on a substrate of oscillating uniform wettability}\label{sec_main}
We consider a droplet on a substrate, where the uniform wettability expressed by the equilibrium contact angle~$\theta_\mathrm{eq}(t)$ oscillates with a frequency~$f$ between a minimum ($\theta_\mathrm{eq}^\mathrm{min}$) and maximum ($\theta_\mathrm{eq}^\mathrm{max}$) value:
\begin{equation}
 \theta_\mathrm{eq}(t) =
 \theta_\mathrm{eq}^\mathrm{min} +
 (\theta_\mathrm{eq}^\mathrm{max} - \theta_\mathrm{eq}^\mathrm{min}) \cdot
 \sin^2 \left(
 \pi f t
\right)
\end{equation}
For further use below, we note that the wettability oscillation is invariant under time reversal, up to a constant phase shift.

We now discuss how the droplet reacts on oscillations in the wettability and compare our numerical results to the outcome from the spherical-cap model.
Then, we show that the induced flow in the droplet is non-reciprocal so that it effectively pumps fluid during one oscillation cycle.

\subsection{Phenomenology}
\label{sec_phenom}

\begin{figure}
 \includegraphics[width=\linewidth]{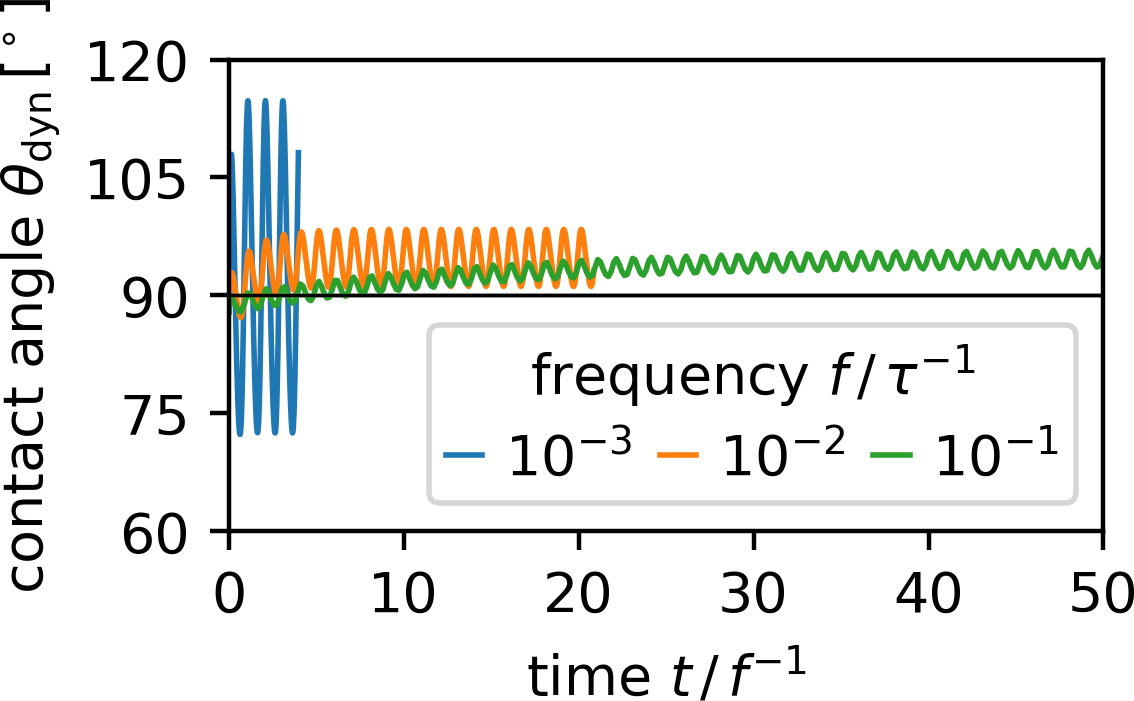}
 \caption{
 Contact angle oscillations in response to oscillations in wettability for three frequencies~$f$  with $\theta_\mathrm{eq}^\mathrm{max} = 120^\circ$ and $\theta_\mathrm{eq}^\mathrm{min}=60^\circ$.
 }
 \label{todo}
\end{figure}

After an initial transient behavior, the droplet settles into a periodic deformation which oscillates with the same frequency as 
the wettability. For three frequencies separated by a factor of 10, we display the dynamic contact angle in Fig.~\ref{todo}.
We observe that the range covered by the dynamic contact angle decreases with increasing frequency.
For slow oscillations the contact angle nearly follows the prescribed equilibrium value of the substrate wettability, while for 
fast oscillations it barely varies leading to an almost steady droplet shape. Note that in our example the droplet does not oscillate 
about the mean value of $\theta_\mathrm{eq}^\mathrm{max} = 120^\circ$ and $\theta_\mathrm{eq}^\mathrm{min}=60^\circ$. 
Furthermore, we observe a phase shift between the oscillating equilibrium and dynamic contact angles, which also depends on $f$.
Note, for different combinations of $\theta_\mathrm{eq}^\mathrm{max}$ and $\theta_\mathrm{eq}^\mathrm{min}$ we provide videos M01--M03 in the Supplementary Material, where a relatively small frequency of $f=10^{-3}\,\tau^{-1}$ is used (for details see Appendix~\ref{appendix_esi}).

To quantify the response of the droplet to the oscillating substrate wettability, we introduce the nonlinear susceptibility~$\chi=|\chi|\exp(\mathrm{i}\Delta\varphi)$ with absolute value~$|\chi|$ and phase shift~$\Delta \varphi$.
The imaginary unit is indicated by~$\mathrm{i}$.
We extract $|\chi|$ from
\begin{equation}
 |\chi(f)|= \frac{\max_t \{\theta_\mathrm{dyn}(t)\} - \min_t \{ \theta_\mathrm{dyn}(t)\}}{\theta_\mathrm{eq}^\mathrm{max} - \theta_\mathrm{eq}^\mathrm{min}}~.
\end{equation}
To calculate the phase shift $\Delta \varphi(f)$, we determine the first Fourier coefficient $\alpha_\mathrm{dyn}$ of the dynamic contact angle $\theta_\mathrm{dyn}$,
\begin{equation}
 \alpha_\mathrm{dyn}(f=T^{-1}) = \lim\limits_{s \to \infty} \frac{1}{T}\int_s^{s+T}\!\!\!\!\!\!\!\!\!\! \theta_\mathrm{dyn}(t)\, \mathrm{e}^{\mathrm{i}\, 2\pi
  t/T} \,\mathrm{d}t \, ,
\end{equation}
where $s \to \infty$ insures that the oscillations of $\theta_\mathrm{dyn}(t)$ are steady.
For the same time intervall we also determine the complex amplitude $\alpha_\mathrm{eq}$ of the prescribed equilibrium contact angle sand then calculate the phase shift between oscillating wettability and dynamic contact angle from
\begin{equation}
\Delta \varphi(f) = \arg[\alpha_\mathrm{dyn}(f)] - \arg[\alpha_\mathrm{eq}(f)] \, ,
\end{equation}
where $\arg$ means the phase angle of the complex amplitude.

\begin{figure}
\includegraphics[width=\linewidth]{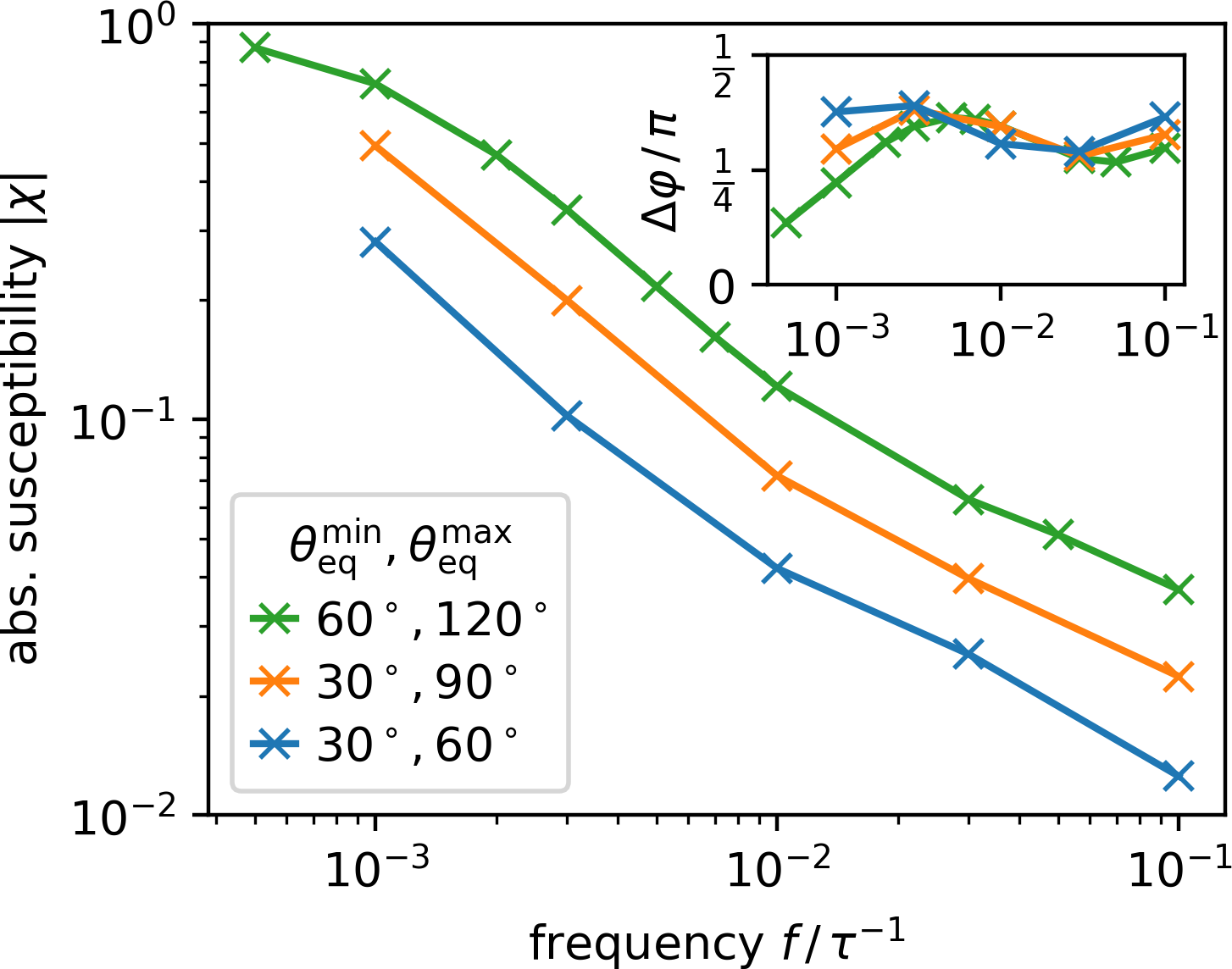}
  \caption{
  Absolute susceptibility $|\chi|$ as a function of oscillation frequency~$f =\omega/2\pi$ in units of $\tau^{-1}$ for three combinations of $\theta_\mathrm{eq}^\mathrm{max}$ and $\theta_\mathrm{eq}^\mathrm{min}$.
  The inset shows the corresponding phase shift $\Delta\varphi$.
  }
 \label{fig_susceptibility}
\end{figure}

In Fig.~\ref{fig_susceptibility} we plot the absolute value $|\chi|$ over frequency for different combinations of $\theta_\mathrm{eq}^\mathrm{max}$ and $\theta_\mathrm{eq}^\mathrm{min}$.
All three curves show the expected decrease of $|\chi|$ with increasing $f$.
Furthermore, for larger difference $\theta_\mathrm{eq}^\mathrm{max} - \theta_\mathrm{eq}^\mathrm{min}$ and larger values of the  equilibrium contact angles, the curves are shifted to larger frequencies but roughly have the same shape.
This suggests by rescaling frequency appropriately, they fall on a master curve.
The inset of Fig.~\ref{fig_susceptibility} plots the corresponding phase shift.
For small frequencies of the oscillating wettability the phase shift tends towards zero for the green curve meaning that the dynamic contact  angle follows the prescribed wetting angle instantaneously.

To gain more insights and also pursue the idea of the master curve further, we compare our observations to the spherical-cap model~\cite{deruijter_contact_1997},  where the droplet always keeps the shape of a spherical cap and the dynamics is solely governed by the Cox-Voinov law for the contact
line.
This will allow us to distinguish phenomena inherent in the contact-line friction from phenomena which are due to the freely deformable liquid-gas interface and the initated fluid flow in the droplet as determined in our boundary element method.
For slow temporal variations in the wettability we expect the spherical-cap model to be valid since fluid flow in the droplet is weak, while for larger frequencies deviations should occur.
We now go into more detail.

In the spherical-cap model the shape of the droplet is constrained to a single degree of freedom for which it determines a dynamic equation~ \cite{deruijter_contact_1997}.
Here, we follow Ref.~\cite{deruijter_contact_1997} and express the model in terms of the dynamic contact angle $\theta_\mathrm{dyn}$,
\begin{equation}
    \frac{\mathrm{d}\theta_\mathrm{dyn}}{\mathrm{d}t} = - g(\theta_\mathrm{dyn}, \theta_\mathrm{eq})
\label{eq_spherical1}
\end{equation}
with
\begin{multline}
g(\theta_\mathrm{dyn}, \theta_\mathrm{eq})  =
 \frac{\gamma_\text{lg}}{9\mu \ln(h/\lambda)}\sqrt[3]{\frac{\pi}{3V}}
 (\theta_\mathrm{dyn}^3 - \theta_\mathrm{eq}^3) \\
\times \sqrt[3]{(1-\cos\theta_\mathrm{dyn})^2\;(2+\cos\theta_\mathrm{dyn})^4}
 \label{eq_spherical2}
\end{multline}
To gain some inside and derive a characteristic relaxation time of the spherical-cap model, we linearize it around $\theta_\mathrm{dyn}=\theta_\mathrm{eq}$ in $\Delta \theta=\theta_\mathrm{dyn}-\theta_\mathrm{eq}$:
\begin{equation}
  \frac{\mathrm{d}\theta_\mathrm{dyn}}{\mathrm{d}t} =  -  \left. \frac{\partial g}{\partial \theta_\mathrm{dyn}}\right|_{\theta_\mathrm{dyn}=\theta_\mathrm{eq}} (\theta_\mathrm{dyn} - \theta_\mathrm{eq}) \,.
\label{eq:linearize}
\end{equation}
The derivative of $g$ is a characteristic relaxation rate~$\tau_0^{-1}$.
But, in our case $\theta_\mathrm{eq}$ is a function of time.
Nevertheless, to have a constant rate, we calculate the derivative at the mean equilibrium contact angle $\bar \theta_\mathrm{eq} = (\theta_\mathrm{eq}^\mathrm{max} - \theta_\mathrm{eq}^\mathrm{min}) /2$ and obtain
\begin{equation}
 \tau_0^{-1}(\bar\theta_\mathrm{eq}) = \frac{\gamma_\text{lg}\bar\theta_\mathrm{eq}^2}{3\mu \ln(h/\lambda)}\sqrt[3]{\frac{\pi}{3V}(1-\cos\bar\theta_\mathrm{eq})^2\;(2+\cos\bar\theta_\mathrm{eq})^4}
\end{equation}
Now, we approximate Eq.~(\ref{eq:linearize}) by using the constant $\tau_0$ instead of the exact derivative of $g$.
Rescaling time by $\tau_0$, we finally arrive at
\begin{equation}
 \frac{\mathrm{d}\theta_\mathrm{dyn}}{\mathrm{d}(t / \tau_0)}
\approx  -[\theta_\mathrm{dyn} - \theta_\mathrm{eq}(t) ]\, ,
  \label{eq_nondim_linear}
\end{equation}
which is the parameter-free linearized model.
Below we will demonstrate that it very nicely fits our computational results for low frequencies and it is the basis for identifying a master curve
for $|\chi( \omega = 2\pi f )|$.

Using time rescaled with $\tau_0$ also in the full spherical-cap model, we can rewrite Eqs.~(\ref{eq_spherical1}) and (\ref{eq_spherical2}) in
a non-dimensionalized form as
\begin{multline}
  \frac{\mathrm{d}\theta_\mathrm{dyn}}{\mathrm{d} (t/\tau_0)} =
 -\frac{1}{3\bar\theta_\mathrm{eq}^2}
 (\theta_\mathrm{dyn}^3 - \theta_\mathrm{eq}(t)^3) \\
\times \sqrt[3]{\frac{(1-\cos\theta_\mathrm{dyn})^2\;(2+\cos\theta_\mathrm{dyn})^4}{(1-\cos\bar\theta_\mathrm{eq})^2\;(2+\cos\bar\theta_\mathrm{eq})^4}}
\label{eq_spherical_nondim}
\end{multline}
Note that here the r.h.s.~is independent of the liquid-gas surface tension~$\gamma_\mathrm{lg}$, the viscosity~$\mu$ of the fluid, and the Cox-Voinov parameter~$\ln(h/\lambda)$, which determines the contact line mobility.
All these parameters are subsumed in the relaxation time~$\tau_0$.
The only remaining parameters are $\theta_\mathrm{eq}^\mathrm{min}$ and $\theta_\mathrm{eq}^\mathrm{max}$, which also determine $\bar \theta_\mathrm{eq}$.

We now explore the linerarized model.
By taking the Fourier transform of Eq.~(\ref{eq_nondim_linear}), one finds the dynamical susceptibility~$\chi(\omega)$, which quantifies how $\theta_\mathrm{dyn}$ responds to an oscillation in $\theta_\mathrm{eq}$:
$\hat\theta_\mathrm{dyn}(\omega) = \chi(\omega) \hat\theta_\mathrm{eq}(\omega)$ with
\begin{equation}
 \chi(\omega) =  \frac{1}{1- \mathrm{i}\,\tau_0\omega} \, .
\label{eq.chi}
\end{equation}
The absolute value~$|\chi|$ reads
\begin{equation}
 |\chi(\omega)| = \sqrt{\frac{1}{1 + (\tau_0\omega)^2}}
 \label{eq_susceptibility}
\end{equation}
and the complex phase~$\varphi$ is
\begin{equation}
 \varphi(\omega) = \arctan(\tau_0 \omega) \, .
 \label{eq_susceptibility2}
\end{equation}
Note that in the region where the linear model applies, $\varphi(\omega)$ is identical to the phase shift $\Delta\varphi (f=\omega/2\pi)$ between $\theta_\mathrm{eq}(t)$ and $\theta_\mathrm{dyn}(t)$ introduced above.

\begin{figure}
\includegraphics[width=0.95\linewidth]{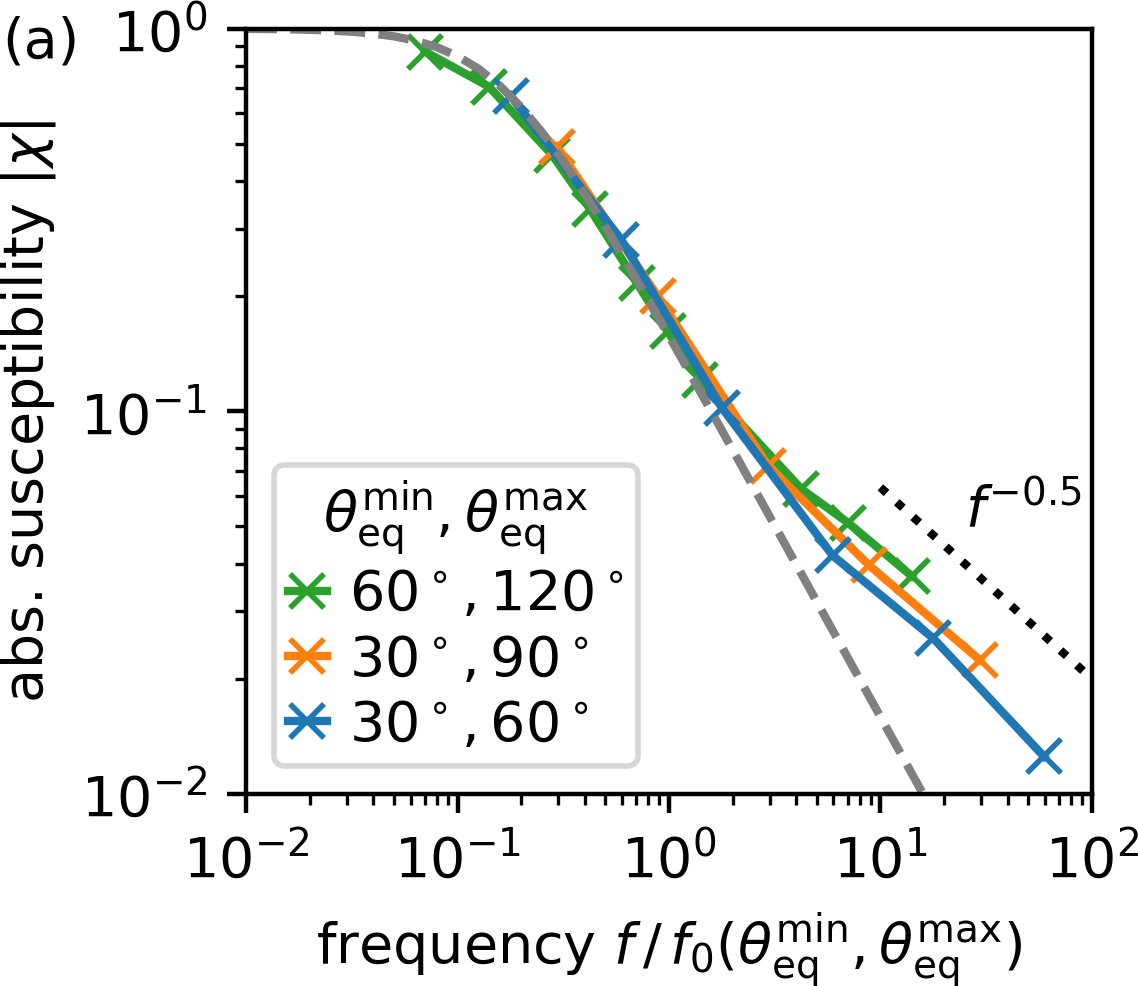}
\includegraphics[width=0.95\linewidth]{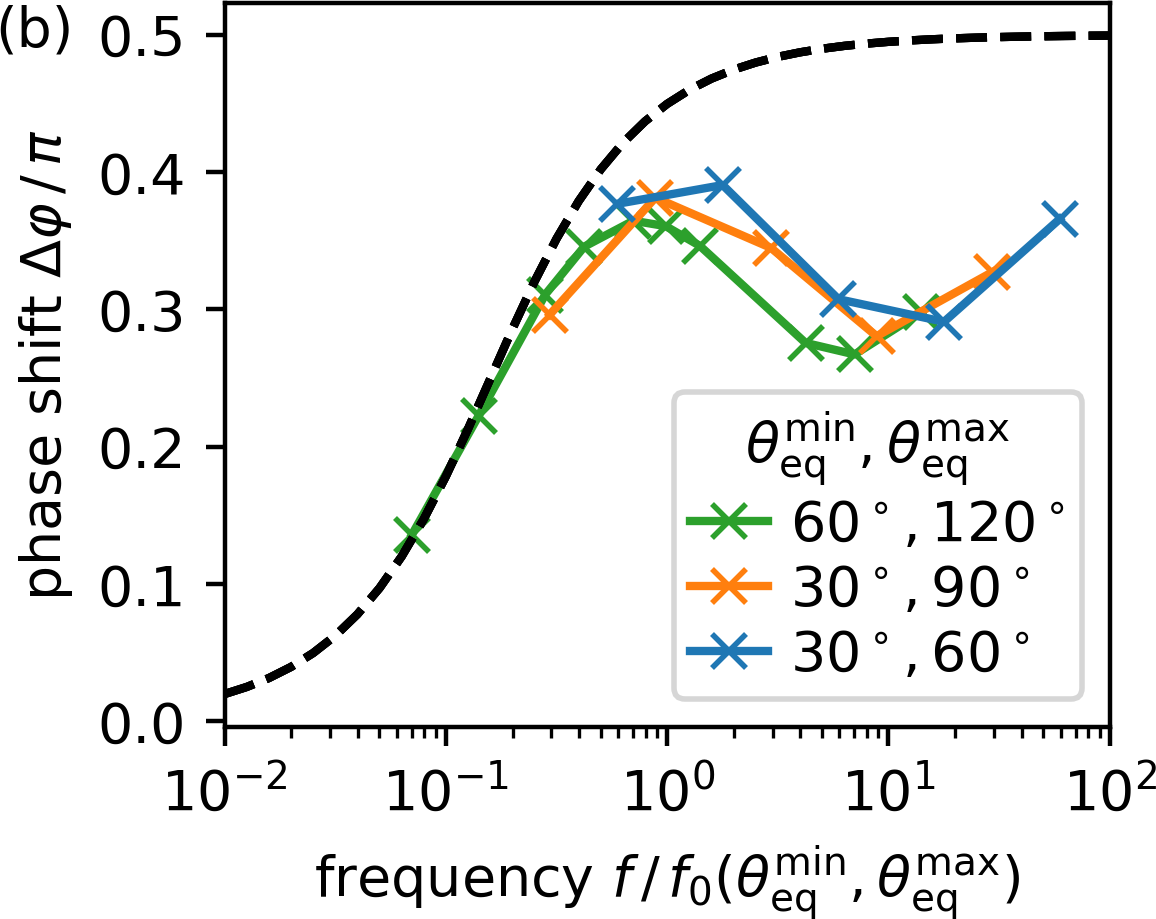}
  \caption{
  Absolute susceptibility $|\chi|$ (a) and phase shift $\Delta\varphi$ (b) as a function of oscillation frequency~$f=\omega/2\pi$ in units of $f_0=\tau_0^{-1}$ determined in the simulations for three combinations of $\theta_\mathrm{eq}^\mathrm{max}$ and $\theta_\mathrm{eq}^\mathrm{min}$ (coloured crosses with solid lines to guide the eye, see legend).
  The dashed grey lines indicate the prediction of the linear model in Eqs.~(\ref{eq_susceptibility}) and (\ref{eq_susceptibility2}), respectively.
  The dotted black line in (a) shows a $f^{-0.5}$ scaling.
  }
 \label{fig_susceptibility_master}
\end{figure}

Because Eqs.~(\ref{eq_susceptibility}) and (\ref{eq_susceptibility2}) do not depend on any material parameters besides the characteristic relaxation time~$\tau_0$, they are candidates for the master curves for our simulation results in Fig.~\ref{fig_susceptibility}.
We simply need to rescale frequency by $\tau_0^{-1}$ calculated for the specific values of $\theta_\mathrm{eq}^\mathrm{min}$ and $\theta_\mathrm{eq}^\mathrm{max}$.
In Figs.~\ref{fig_susceptibility_master}(a) and (b) we display the master curves as dashed lines together with the results from our boundary element method (BEM) using rescaled frequencies.
First, in Fig.~\ref{fig_susceptibility_master}(a) we observe that the BEM results all fall on a common master curve for $f< \tau_0^{-1}$, which perfectly matches the linear model in this range.
For $ f > \tau_0^{-1}$ the BEM results do not follow a common master curve and they deviate from the linear model.
We observe that the absolute susceptibility enters an algebraic decay with an approximate exponent $-0.5$ rather than $-1$ as predicted by the linear model.
Second, in Fig.~\ref{fig_susceptibility_master}(b) we similarly observe that the BEM results approach the linear model for small $f$, however they start to deviate siginificantly for $f \tau_0 > 3 \cdot 10^{-1}$.
The phase shift varies between roughly $0.3\pi$ and $0.4\pi$ rather than approaching $0.5\pi$ as predicted analytically by the linear model in Eq.~(\ref{eq_susceptibility2}) and indicated as dashed line.
The phase shifts beyond $f \tau_0 > 3 \cdot 10^{-1}$ apparently depend on $\theta_\mathrm{eq}^\mathrm{max}$ and $\theta_\mathrm{eq}^\mathrm{min}$.

To interpret these observations, we distinguish two regimes: a low frequency regime with $f\tau_0 < 1$ and a high frequency regime with $f\tau_0 > 1 $.
In the low frequency regime, the oscillations are sufficiently slow so that the droplet can adapt its shape and keep it close to a spherical cap.
Thus, the linear spherical-cap model is valid.
In fact, in the limit of vanishing $f$ it becomes exact as the dynamics becomes quasistatic.
According to linear response theory, the imaginary part $\mathrm{Im}\chi$, which in our linear model from Eq.~(\ref{eq.chi}) reads
\begin{equation}
\mathrm{Im} \chi(\omega) =  \frac{\tau_0 \omega}{1 + (\tau_0\omega)^2} \, ,
\label{eq.Imchi}
\end{equation}
quantifies dissipation.
Because $\mathrm{Im}\chi$ as well as the phase shift $\Delta\phi$ from Eq.~(\ref{eq_susceptibility2}) are linear in $\omega$ at small frequencies, this implies that almost no work performed on the droplet through slow wettability oscillations is dissipated by the friction of the contact line.

In the high frequency regime, the droplet deviates far from the linear model since small $|\chi|$ means $\theta_\mathrm{dyn}$ barely tracks $\theta_\mathrm{eq}(t)$.
But also the full spherical cap model is unable to predict the observed behavior for $|\chi|$ and $\Delta \varphi$ and, in particular, the deviation from a common master curve.
This implies that the droplet shape deviates from a spherical cap when fast wettability oscillations are applied.
These deviations occur close to the contact line, which can move quickly without displacing much liquid and thereby bends the free surface, \emph{i.e.}, it locally in- or decreases curvature, which is precluded in the spherical cap model.
This explains why $\theta_\mathrm{dyn}$ is more susceptible at large $f$, meaning it lies above the prediction of the linear model:
A small adjustment of the position of the contact line can drastically alter $\theta_\mathrm{dyn}$ and increase in a short time the surface energy of the droplet relative to the equilibrium reference shape for a given $\theta_\mathrm{eq}$.
Thus the work performed on the droplet is not completely dissipated.
Therefore, the phase shift angle $\Delta \varphi$ is below the value $\pi/2$, which is expected for a complete dissipation and which is predicted within the spherical-cap model for large frequencies.

The spherical cap model only includes dissipation at the contact line.
It does not account for viscous friction, friction at the substrate interface, or the elasticity of the free surface.
All of these are present in the BEM results, however.
We now want to focus on these contributions by turning our attention toward the internal flow of the droplet and toward its deformation from the spherical cap shape.

\subsection{Deformation and pumping} \label{sec_pumping}
The shape oscillations of the droplet are accompanied by internal fluid flow.
When the wettability increases, the droplet wets more area on the substrate and fluid moves from the top of the droplet through its center to the contact line.
When the wettability decreases, the droplet wets less area on the substrate and fluid moves in the opposite direction.

\begin{figure}
 \includegraphics[width=\linewidth]{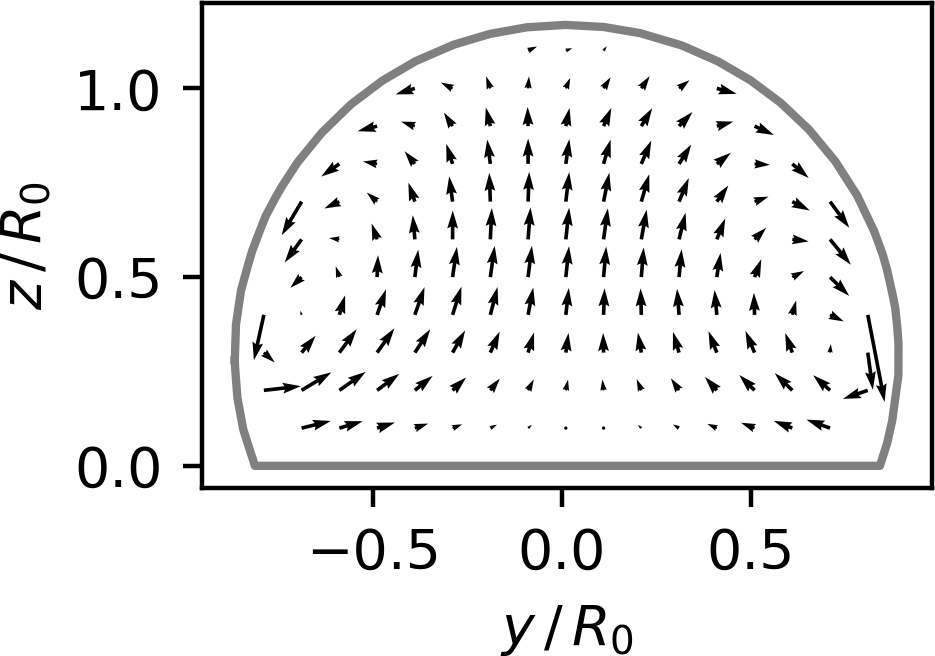}
 \caption{Displacements $\vec d(\vec r)$ of tracers after one full period in a cross section of a droplet on a substrate with oscillating wettability with $\theta_\mathrm{eq}^\mathrm{min}=60^\circ$, $\theta_\mathrm{eq}^\mathrm{max}=120^\circ$, and $f=10^{-3}\,\tau^{-1}$.}
 \label{figure10}
\end{figure}

However, this back-and-forth does not cancel out completely and there is a net displacement of fluid after each period, \emph{i.e.}, fluid is pumped within the droplet, as illustrated in Fig.~\ref{figure10}.
To quantify the net displacement, we place point-like tracer particles in the droplet and track their motion for one full period of oscillation, so that to each starting point $\vec r_0$ we can assign a displacement
\begin{equation}
 \vec d (\vec r_0) =  \vec r(T) - \vec r_0 \,  ,
\end{equation}
where $\vec r(t)$ is the solution of the differential equation
$\dot{\vec r}(t) = \vec v( \vec r,t)$
with initial condition $\vec r(0) = \vec r_0$ and $\vec v(\vec r, t)$ is the interior velocity field of the droplet when it is steadily oscillating.
In Fig.~\ref{figure10} we display an example of $\vec d (\vec r)$ that is representative for all studied cases.
Qualitatively, it shows a circulation of fluid inside the droplet with fluid travelling up through the center of the droplet and down along its free surface in a single toroidal vortex, which covers the interior of the droplet completely.
We return to this observation after considering the shape of the droplet.

\begin{figure}
 \includegraphics[width=\linewidth]{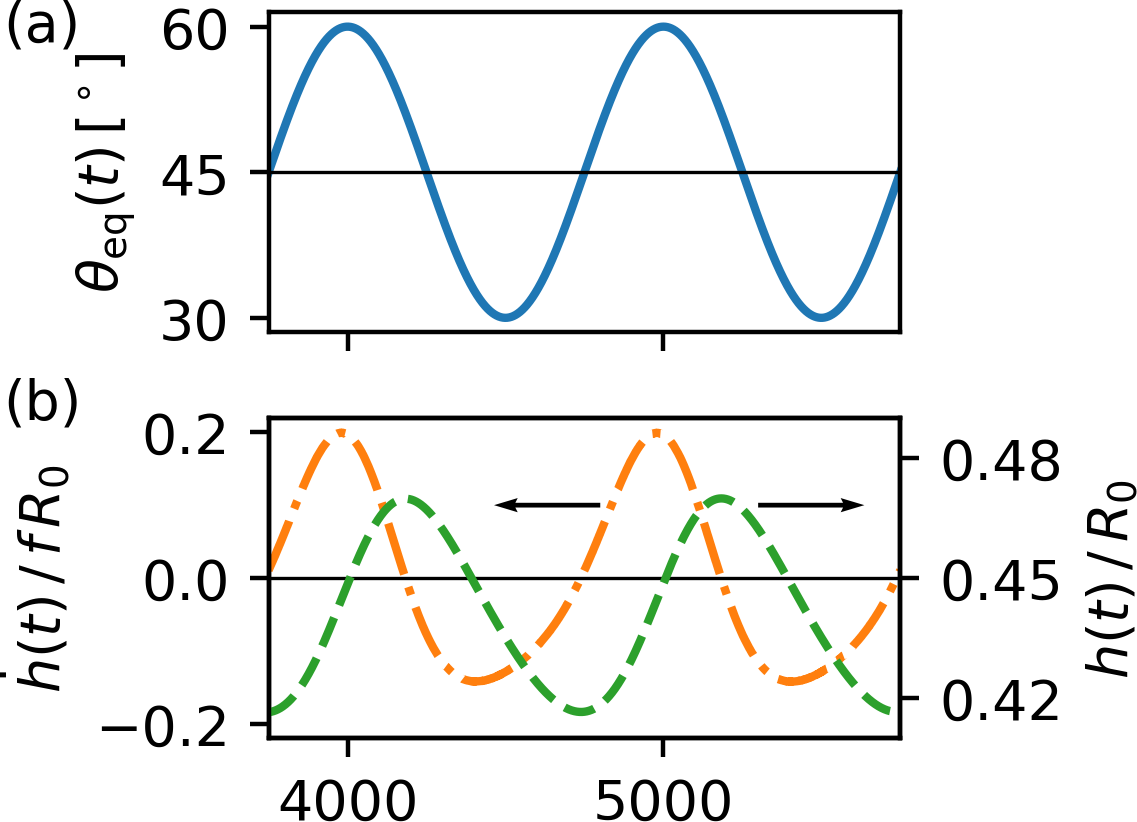}
 \caption{Droplet shape changes for $f=10^{-3}\,\tau^{-1}$, $\theta_\mathrm{eq}^\mathrm{max}=60^\circ$ and $\theta_\mathrm{eq}^\mathrm{min}=30^\circ$.
 (a)~Wettability of the substrate as a function of time characterized by the equilibrium contact angle~$\theta_\mathrm{eq}$.
(b) Deformation of the droplet quantified by the height~$h(t)$ and rate of deformation quantified by $\dot h(t)$ plotted versus time.
Note that $\dot h(t)$ indicates unequal upward and downward slopes for  the height dynamics.
}
 \label{figure7}
\end{figure}

To quantify changes in the droplet shape, one can use, \emph{e.g.}, the dynamic contact angle~$\theta_\mathrm{dyn}$ or the droplet height~$h$.
In Fig.~\ref{figure7} we plot two oscillation cycles of $\theta_{\text{eq}}(t)$ in graph~(a) together with the corresponding $h(t)$ and its time derivative $\dot{h}(t)$ in graph~(b).
We first observe that the height of the droplet adjusts faster to a decrease in wettability than to an increase;
meaning, the upward slope of $h(t)$ corresponding to an increase of $\theta_{\text{eq}}(t)$ is larger than the magnitude of the downward slope.
We understand this by studying the mobility of the contact line as a function of the dynamic contact angle.
When the droplet is equilibrated, \emph{i.e.}, $\theta_\mathrm{eq}=\theta_\mathrm{dyn}$ and a small change in wettability occurs, \emph{i.e.} 
$\theta_\mathrm{eq} \to \theta_\mathrm{eq} + \delta\theta_\mathrm{eq}$,
we calculate the contact line mobility $m$ by linearizing the Cox-Voinov law, Eq.~(\ref{eq_coxvoinov}).
The velocity of the contact line becomes
\begin{equation}
 v_\mathrm{contact} \approx - m  \delta\theta_\mathrm{eq} \quad \text{with} \quad
 m = \frac{\gamma \theta_\mathrm{eq}^2 }{3\mu \ln (h/\lambda)} \,
\end{equation}
Apparently, the mobility~$m$ can be smaller or larger for droplets with the same $\theta_\mathrm{dyn}$, depending on $\theta_\mathrm{eq}$.
If the wettability is increasing (smaller $\theta_\mathrm{eq}$) $m$ is decreased.
Respectively, if the wettability is decreasing (larger $\theta_\mathrm{eq}$) $m$ is increased.
Thus, when wettability increases over time, contact line mobility
decreases and the droplet's shape adjusts more slowly than when wettability decreases over time.
This explains the behavior of $h(t)$ in response to $\theta_{\text{eq}}(t)$.

However, varying rates of deformation are insufficient to explain a net pumping of the liquid because the equations of Stokes flow, Eq.~(\ref{eq_stokes}), are independent of time.
Therefore, a different aspect of the deformation must be responsible for the pumping.

\begin{figure}
 \includegraphics[width=\linewidth]{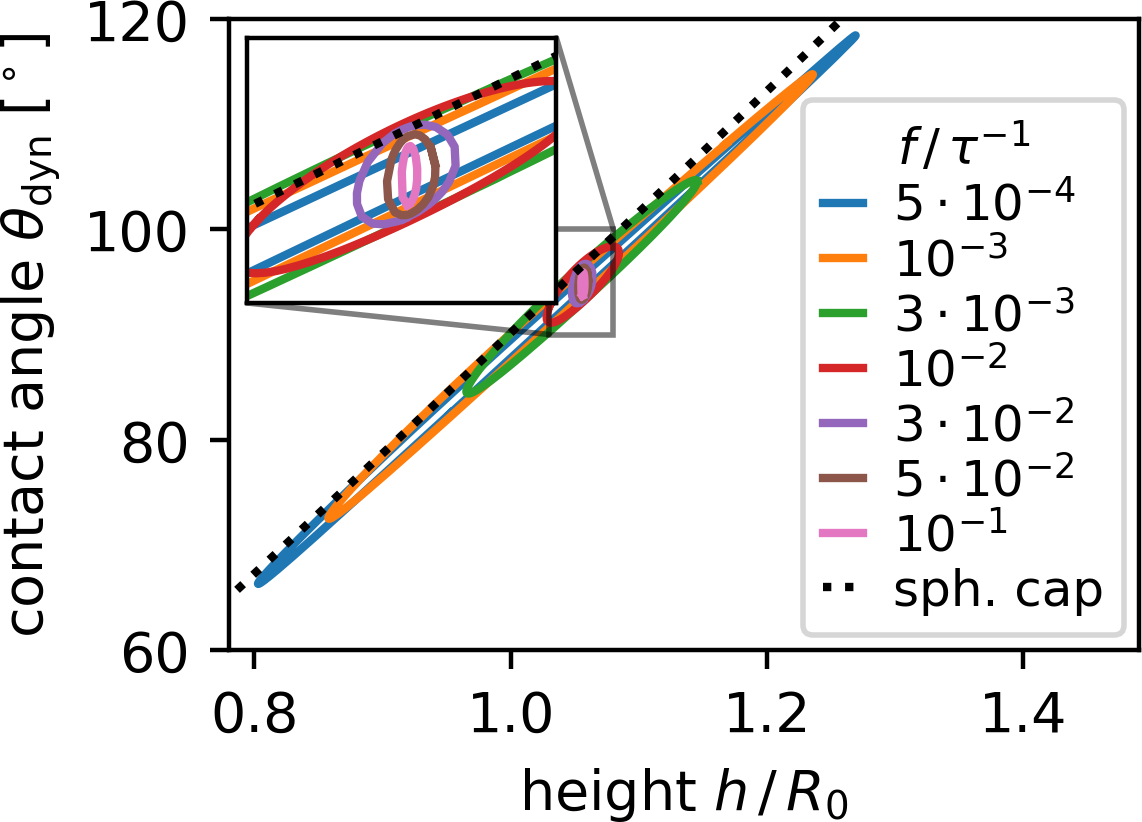}
 \caption{
Closed trajectories in configuration space projected onto the $h$-$\theta_\mathrm{dyn}$ plane for various frequencies~$f$ (see legend) for $\theta_\mathrm{eq}^\mathrm{min}=60^\circ$ and $\theta_\mathrm{eq}^\mathrm{max}=120^\circ$.
 The limiting behavior of the spherical-cap model with $f \tau_0 \ll 1$ is indicated by the black dashed line.}
 \label{figure2}
\end{figure}

Unlike the spherical cap model our BEM simulations are not constrained to a single degree of freedom.
So as a minimal extension for describing the temporal shape variations of the droplet, we investigate the coupled dynamics of \emph{two} degrees of freedom, contact angle~$\theta_\mathrm{dyn}$ and droplet height~$h$.
In Fig.~\ref{figure2} we represent the droplet dynamics in the configuration space spanned by these two variables; one oscillation of the droplet corresponds to a closed trajectory.
Interestingly, the non-zero area enclosed by the trajectory reveals the dynamics of the droplet as non-reciprocal, meaning under time-reversal the dynamics looks different.
In our concrete case the droplet assumes slightly different shapes during increasing and decreasing wettability in the course of one period.
The non-reciprocal dynamics is clearly due to the flow field generated inside the droplet.
Since the spherical-cap model only has one dynamic variable, its dynamics can only be reciprocal.
The dashed line in Fig.~\ref{figure2} shows the model prediction for vanishing frequency.%
\footnote{Note that due to the discretization of the droplet surface, all $\theta_\mathrm{dyn}$ of the simulated curves are systematically
shifted downwards by a small angle.
Otherwise, they should
center on the dashed line.}
Note the non-zero area also means that contact angle and height oscillate out-of-phase with each other.
From the study of microswimmers at vanishing Reynolds numbers, we know
Purcell's scallop theorem~\cite{purcell_life_1977} which states
that non-reciprocal shape changes are needed for microswimmers to move forward.
Similarly, the pumping displacement mentioned in the beginning is linked to the non-reciprocal droplet deformation.

To quantify the non-reciprocal shape dynamics of the droplet, we take the area enclosed by the trajectory in configuration space,
which, in general, is high-dimensional.
However, projecting this trajectory into the two-dimensional space spanned by droplet height~$h$ and contact angle~$\theta_\mathrm{dyn}$, we can immediately calculate the projected area as
\begin{equation}
 A = \lim\limits_{s\to\infty}
 \oint_s^{s+T} \!\!\!\!\!\!\!\!\!\!\! h(t) \, \dot\theta_\mathrm{dyn}(t)\,\mathrm{d}t
\end{equation}
where the limit $s\to\infty$ ensures that $h(t)$ and $\theta_\mathrm{dyn}(t)$ perform steady oscillations.
In the following, we call the parameter $A$ shape non-reciprocity.

\begin{figure}
\raggedleft
 \includegraphics[width=\linewidth]{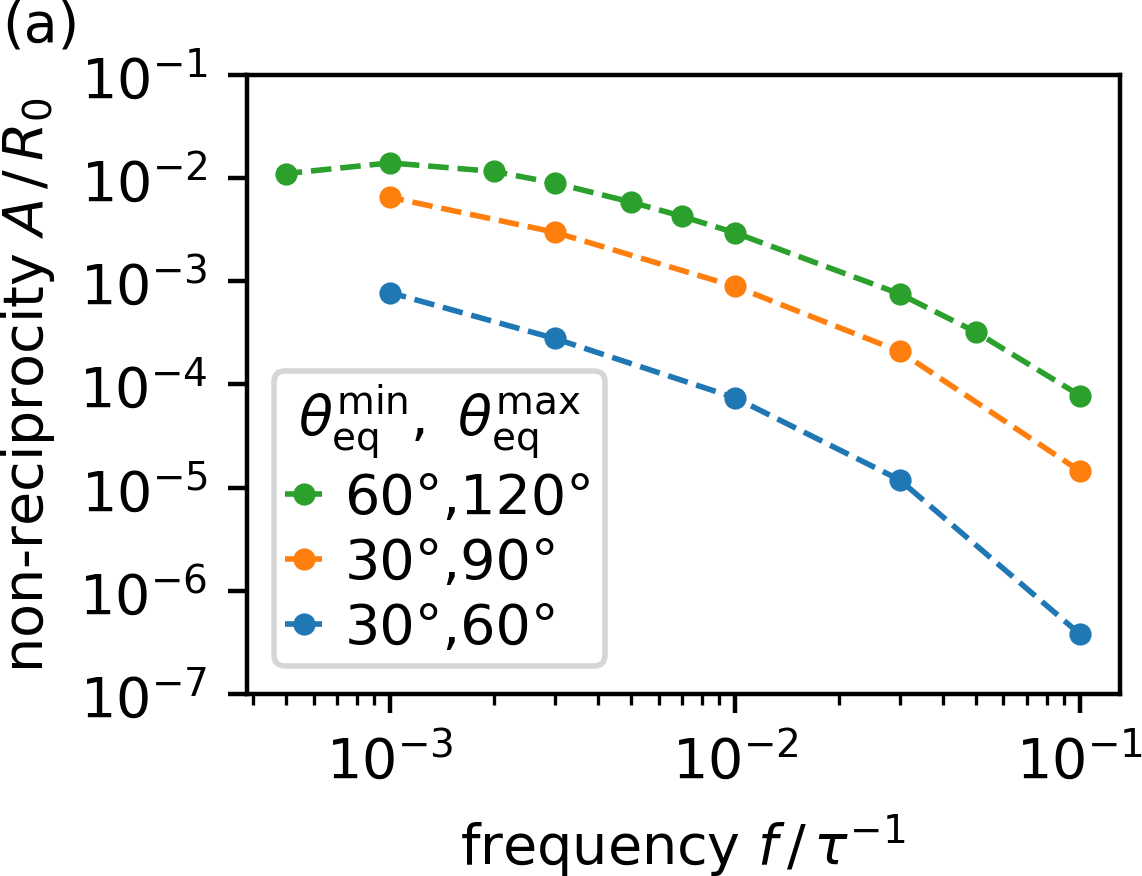}
 \includegraphics[width=0.935\linewidth]{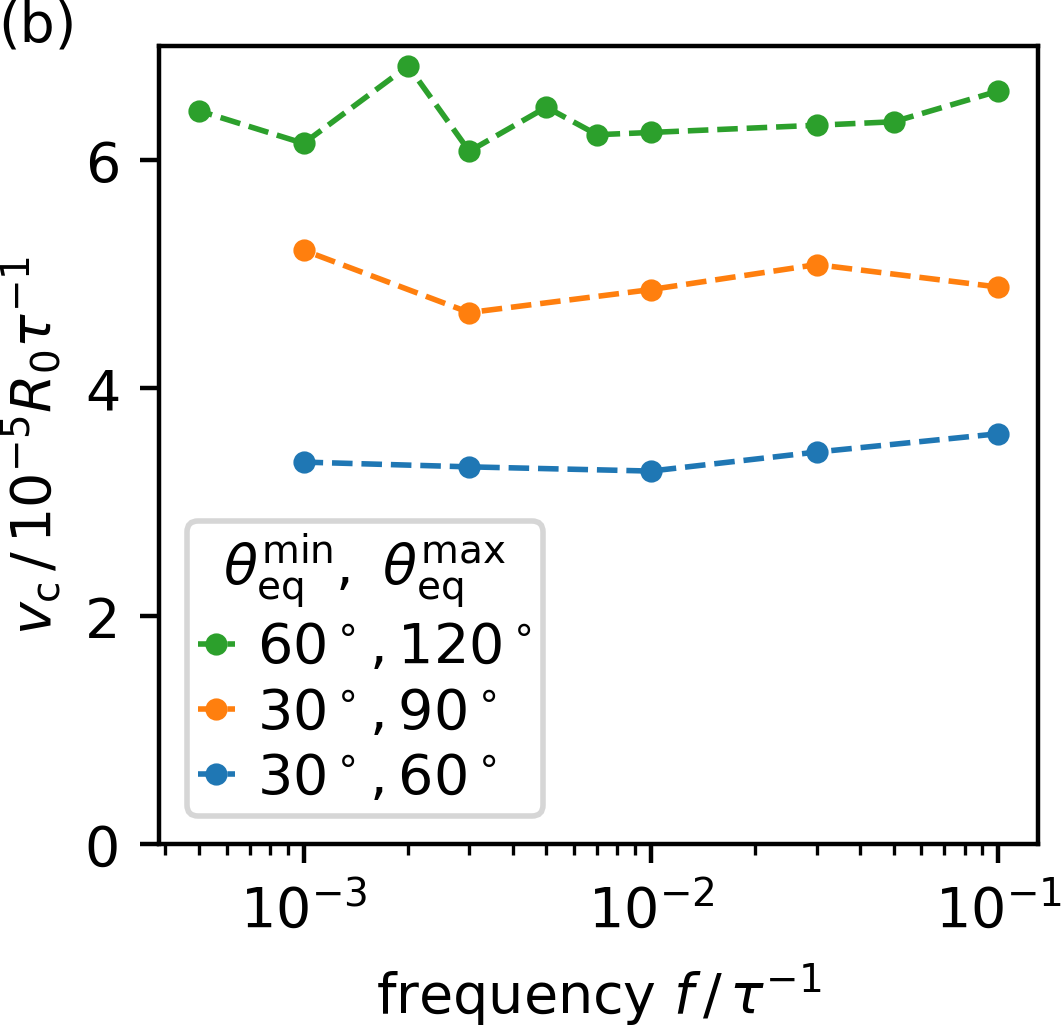}
 \includegraphics[width=\linewidth]{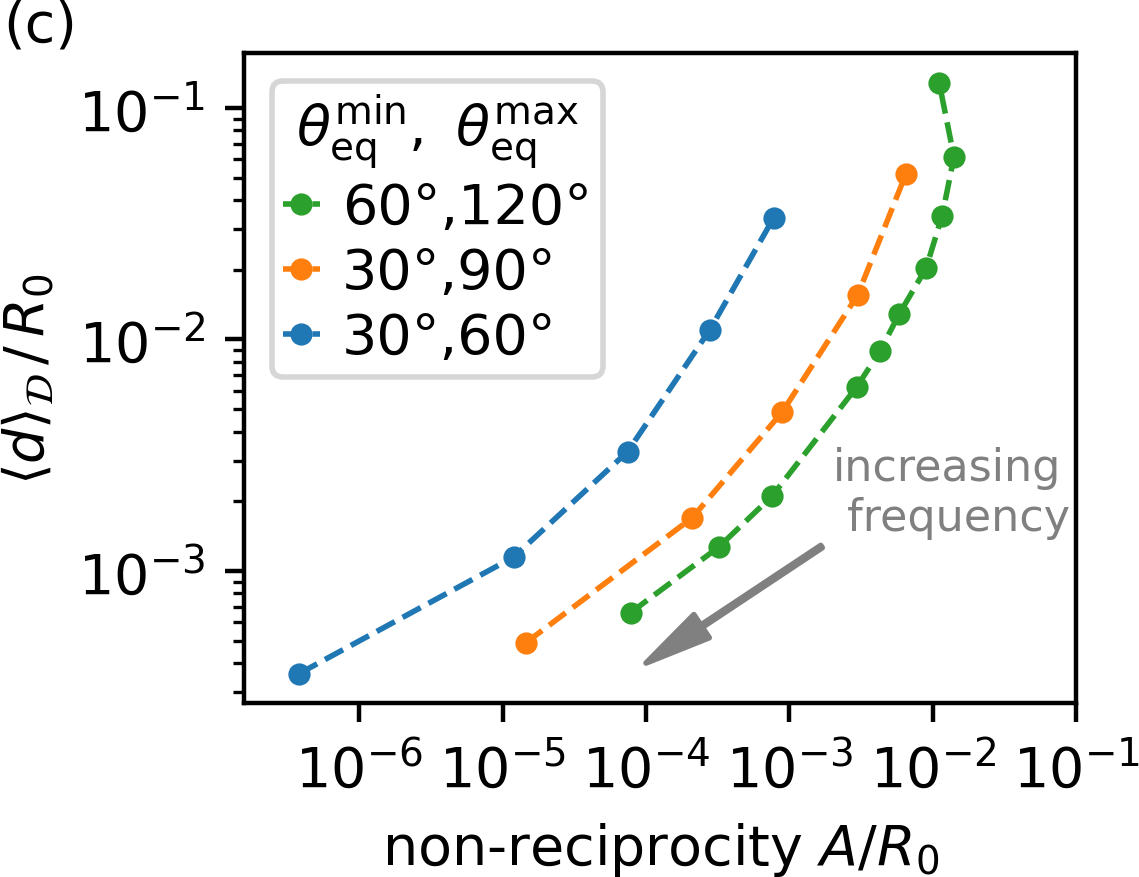}
 \caption{
 Shape non-reciprocity~$A$~(a) and pumping velocity~$v_\mathrm{c}=f\langle d\rangle_\mathcal{D}$ (b) as a function of oscillation frequency~$f$ determined for the three combinations of $\theta_\mathrm{eq}^\mathrm{min}$ and $\theta_\mathrm{eq}^\mathrm{max}$ (colored points with dashed lines to guide the eye, see legend).
 (c)~Displacement $\langle d \rangle_\mathcal{D}$ plotted directly against~$A$ for the same combinations, with an arrow indicating the direction of increasing~$f$.
 }
 \label{figure3}
\end{figure}

The closed trajectories or loops in the configuration space presented in Figure \ref{figure2} extend further in both dimensions and enclose a larger area~$A$ as frequency decreases.
An increasing area~$A$ means the dynamics of the droplet shape becomes less reciprocal.
Figure~\ref{figure3}(a) displays the dependence of $A$ on frequency~$f$ for all three cases of equilibrium contact angles.
We always observe that $A$ decreases for large $f$, while in the case of $\theta_\mathrm{eq}^\mathrm{min}=60^\circ$ and $\theta_\mathrm{eq}^\mathrm{max}=120^\circ$ the non-reciprocity $A$ has a maximum and decreases toward small~$f$.
The latter is expected since in the quasistationary case of vanishing $f$ the motion has to become reciprocal.
Notably, even at small frequencies $A$ is relatively large which means even though the dynamic response approaches that of the spherical cap model, there is still a significant difference between our BEM dynamics and the spherical cap model.

We can directly understand the non-zero values for $A$ by considering the basic mechanisms driving the droplet.
When wettability changes, it gives rise to uncompensated Young forces~\cite{young_essay_1805} at the contact line.
The contact line starts to move, which brings $\theta_\mathrm{dyn}$ closer to $\theta_\mathrm{eq}$ and relaxes the Young forces.
At the same time, the free surface is bent locally thereby introducing uncompensated surface stresses in the vicinity of the contact line.
Those stresses redistribute liquid inside the droplet and eventually affect droplet height~$h$.
So, while $\theta_\mathrm{dyn}$ adjusts directly to the changes in wettability, the effect on~$h$ is mediated by the initiated flow and, therefore, delayed.

In the quasistatic limit, $f\to0$, that delay becomes negligible relative to $f$; contact angle and height oscillate in synchrony and the droplet's behavior approaches the spherical cap model, Eq.~(\ref{eq_spherical_nondim}).
In the limit $f\to\infty$ the surface stresses relax not by redistributing liquid, but because wettability quickly returns to its original value, thereby eliminating the uncompensated Young forces before the contact line moves significantly.
Here, the droplet hardly oscillates and thereby approaches a static spherical cap shape.
So, in both limits the droplet assumes spherical shapes and only intermediate frequency values~$f$ cause a significant deviation from this form.

We now investigate the liquid displacement $\vec d (\vec r_0)$ inside the droplet in more depth.
To show the link between non-reciprocity and displacement quantitatively, we consider the median pumping speed $v_c=f\langle d \rangle_\mathcal{D}$ w.r.t.~the interior of the droplet, \emph{i.e.}, the 50th percentile value of the spatial distribution of $d = |\vec d(\vec r_0)|$ multiplied by frequency~$f$.
In Fig.~\ref{figure3}(b) we observe that $v_c$ remains constant for all $f$ with some fluctuations which means it is purely determined by the material properties of the liquid and substrate and not the oscillation frequency.
We already mentioned since the spherical-cap model only has a single degree of freedom, it cannot exhibit any circulatory pumping according to the scallop theorem.
So, just like the non-reciprocity~$A$, the non-zero median pumping speed~$v_\mathrm{c}$ also shows that the spherical cap model cannot completely describe the droplet's behavior for small $f$.
However, we expect that $v_\mathrm{c}$ eventually tends to zero since for sufficiently slow motion of the contact line, the droplet will go through a sequence of spherical-cap shapes.
Constraints in the simulation time make it impossible to reach the limit $f \rightarrow 0$.
However, we observe in Fig.~\ref{figure2} and the inset that for $f < 10^{-2}\,\tau^{-1}$ the distance between the two halves of the closed trajectories decreases with decreasing $f$ such that the area $A$ and therefore also $v_\mathrm{c}$ should ultimately vanish.
Note, an analogous behavior was observed, for example, in simulations of the one-armed microswimmer \cite{gauger_numerical_2006}.
When its flexible flagellum beats quickly, it bends due to frictional forces and thereby moves nonreciprocally.
But when it beats very slowly, it behaves like a rigid rod and thus moves reciprocally and the microswimmer cannot swim forward.

Finally, in Fig.~\ref{figure3}(c) we relate the median displacement~$\langle d \rangle_\mathcal{D}$ directly to the non-reciprocity $A$ and identify a non-linear relation.
We checked that the scaling from Sec.~\ref{sec_phenom}, derived from the spherical cap model, does not produce a master curve for either $A$ or $\langle |\vec d|\rangle_\mathcal{D}$.
This corroborates further that the spherical cap model is inapplicable for these quantities.

\subsection{Oscillations of nonuniform wettability profiles}
\label{sec_nonuniform}
\begin{figure}
  \includegraphics[width=\linewidth]{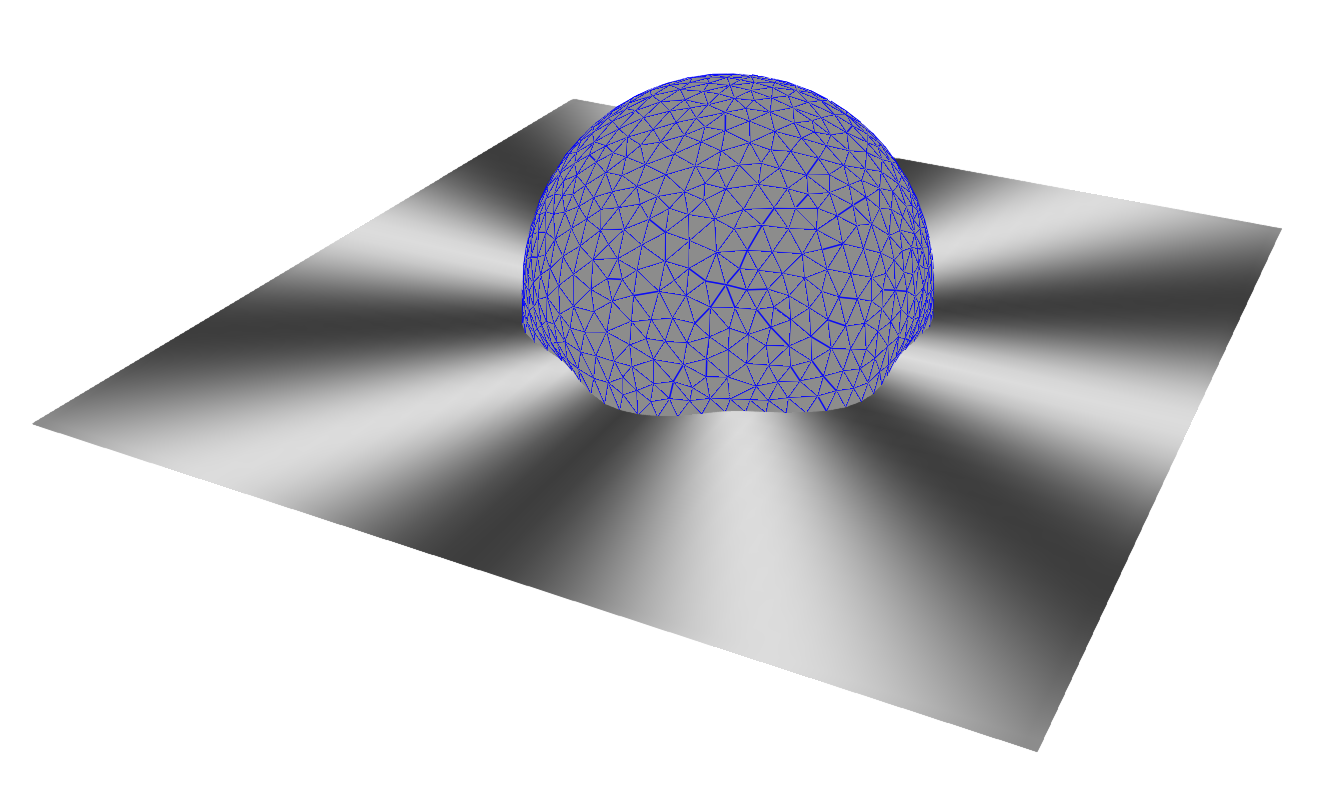}
  \caption{%
    Snapshot of a droplet on a substrate with a nonuniform wettability pattern with six-fold symmetry.
    The equilibrium contact angles varies between $\theta_\mathrm{eq}^\mathrm{min}=60^\circ$ and $\theta_\mathrm{eq}^\mathrm{max}=120^\circ$.
  }
  \label{fig_windmill_3}
\end{figure}

We now extend our investigation to oscillating nonuniform patterns of wettability.
Specifically, we choose a pattern where the wettability varies periodically along the contact line of the droplet.
As an example, we take the six-fold pattern illustrated in Fig.~\ref{fig_windmill_3}, where the equilibrium contact angle is modulated in space and time:
\begin{multline}
 \theta_\mathrm{eq}(\vec r, t) =
 \theta_\mathrm{eq}^\mathrm{max}
 -(\theta_\mathrm{eq}^\mathrm{max} - \theta_\mathrm{eq}^\mathrm{min}) \cdot
 \sin^2\left(\pi f t\right) \cdot \\
 \sin^2 \left(\frac{n}{2} \phi \right)
 \left[1-\exp{\left(-10\frac{|\vec r|^2}{R_0^2}\right)}\right]
 \label{eq_nonuniform}
\end{multline}
with $\vec r = (r \cos \phi, r \sin \phi, 0)^\mathrm{T}$ and $n=6$.
The droplet initially sits with its base area centered at $\vec r=0$.
In an experiment, an equivalent light intensity pattern can be realized with  Laguerre-Gauss laser modes~\cite{allen_orbital_1992}.

After an initial transient behavior, the droplet settles into steady oscillations analogous to the initial behavior in Sect.~\ref{sec_phenom}.
However, in this case the droplet shape is not axisymmetric but instead it follows the six-fold symmetry of the wettability pattern.
A snapshot of the droplet shape is presented in Fig.~\ref{fig_windmill_3} and the whole dynamics can be seen in a video in Supplementary Material~M04.

\begin{figure}
  \includegraphics[width=\linewidth]{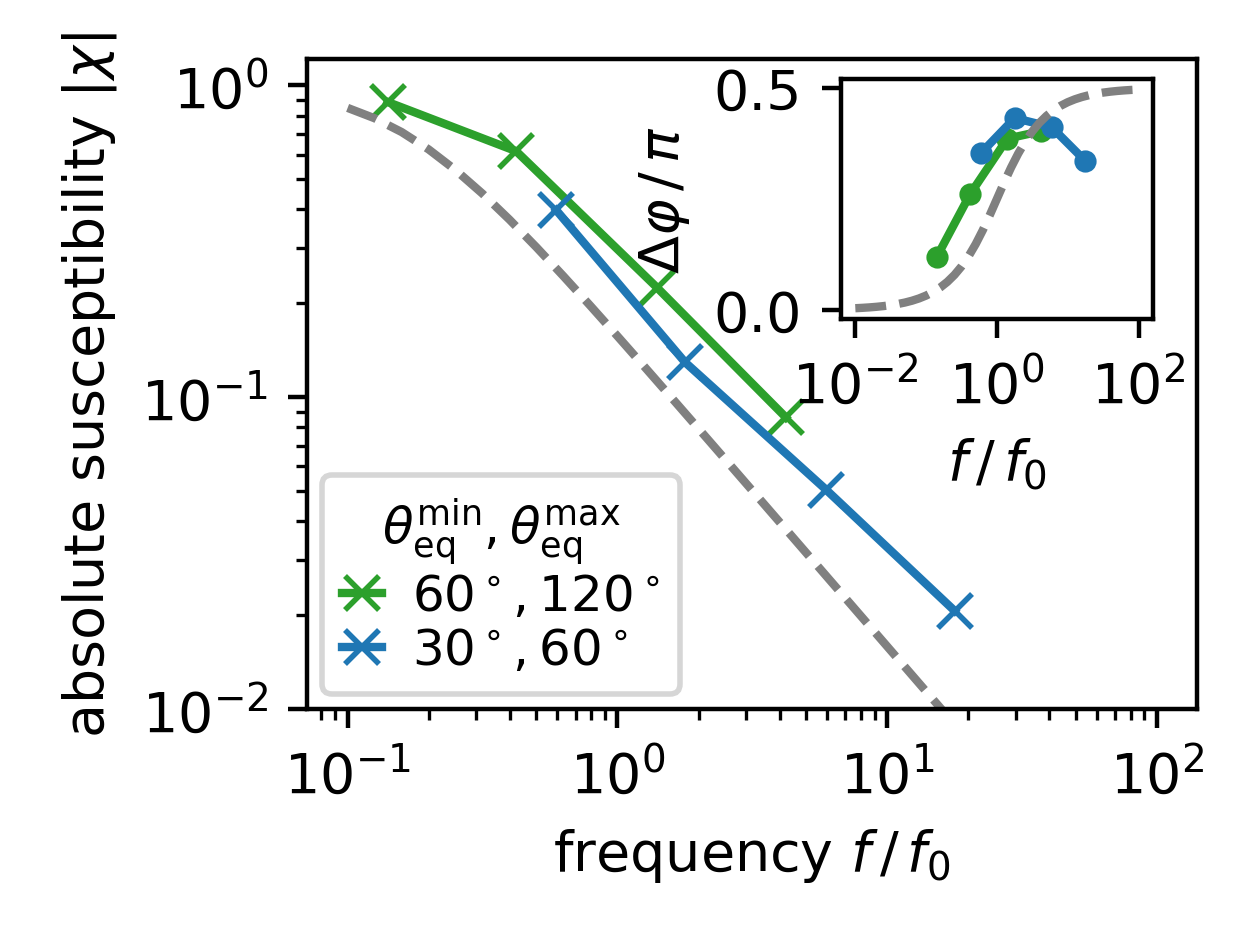}
  \caption{
  Absolute susceptibility $|\chi|$ and phase shift~$\Delta \varphi$ (inset) as a function of oscillation frequency~$f$ in units of $f_0=\tau_0^{-1}$ determined in the simulations for two combinations of $\theta_\mathrm{eq}^\mathrm{max}$ and $\theta_\mathrm{eq}^\mathrm{min}$ (coloured crosses and dots with solid lines to guide the eye, see legend). The dashed grey lines indicate the  prediction of the linear model in Eqs.~(\ref{eq_susceptibility}) and (\ref{eq_susceptibility2}), respectively.
  }
  \label{fig_windmill_1}
\end{figure}

To study the steady oscillations quantitatively, we introduce the 6th harmonic mode~$a_6$ of the dynamic contact angle along the contact line,
\begin{equation}
a_6 = \frac{1}{L}\int_{0}^{L} \theta_\mathrm{dyn}(s)\cos(\mathrm{i}\,k_6 s) \,\mathrm{d}s
\end{equation}
with $k_6=6 \cdot 2\pi/L$, the instantaneous length~$L$ of the contact line, and the dynamic contact angle $\theta_\mathrm{dyn}(s)$ parameterized by the arc length~$s$ of the contact line.
In the same way, we calculate the 6th harmonic mode~$\tilde a_6$ of the equilibrium contact angle~$\theta_\mathrm{eq}(s)$.
Both, $a_6$ and $\tilde a_6$, are periodic functions in time~$t$ when the droplet has reached steady oscillation and so the ratio of their Fourier transforms in time gives again a susceptibility~$\chi(f)$.
In Fig.~\ref{fig_windmill_1} we display the absolute value~$|\chi(f)|$ and the phase shift~$\Delta \varphi$ given by $\chi=|\chi|\mathrm{e}^{\Delta\varphi}$.
The principal behavior is similar to that of an oscillating uniform pattern shown in Fig.~\ref{fig_susceptibility_master}.
However, now the linear model derived from the spherical cap model does not provide a master curve anymore since the droplet shape deviates strongly from the spherical cap.
In detail, we observe while $|\chi|$ is shifted toward larger~$f$ when compared to the linear model (dashed line), $\Delta\varphi$ is shifted toward smaller $f$.

\begin{figure}
  \includegraphics[width=\linewidth]{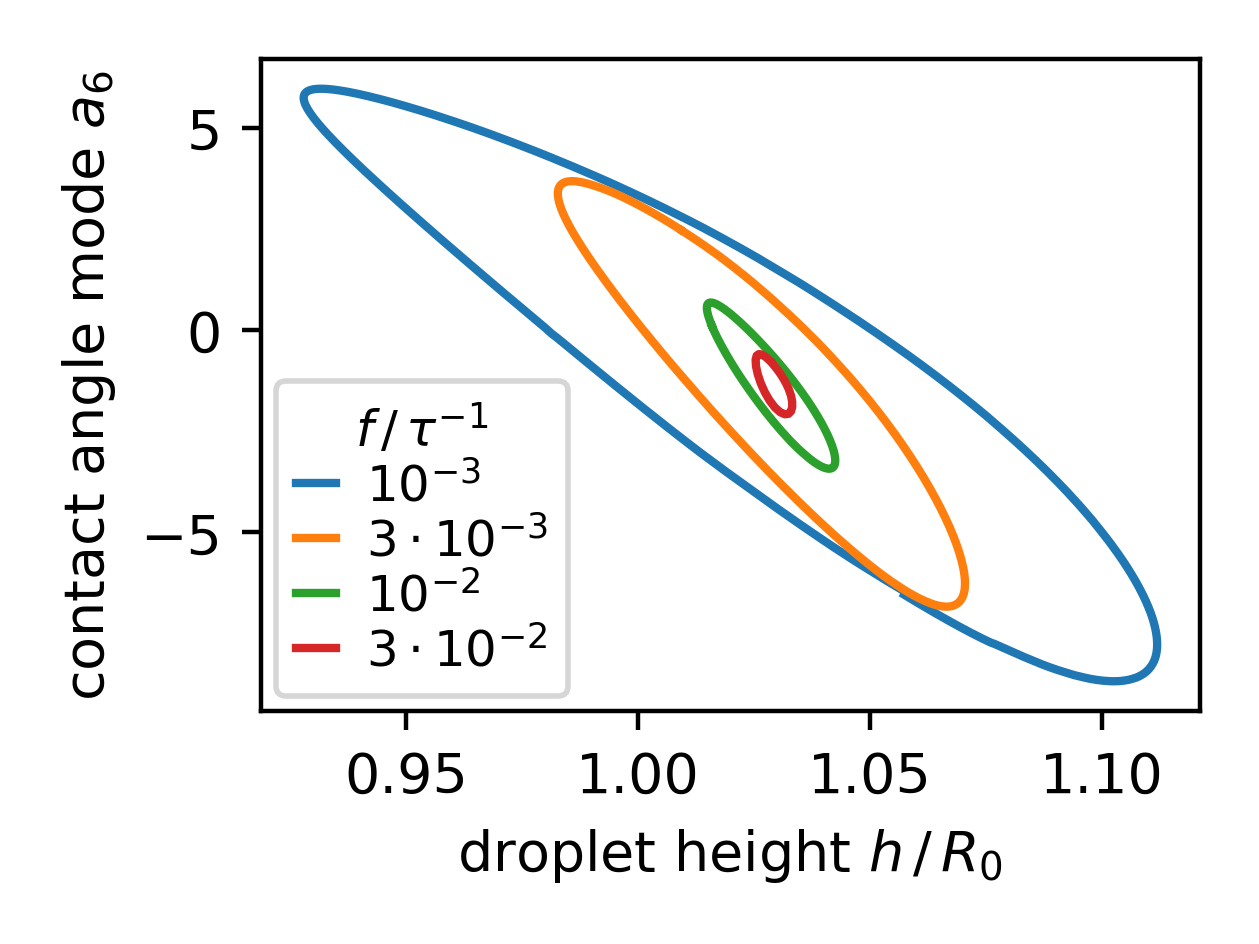}
  \caption{Closed trajectories in configuration space projected onto the $h$-$a_6$ plane for various frequencies~$f$ (see legend) for $\theta_\mathrm{eq}^\mathrm{min}=60^\circ$ and $\theta_\mathrm{eq}^\mathrm{max}=120^\circ$.}
  \label{fig_windmill_2}
\end{figure}

As before in Sect.~\ref{sec_pumping}, we study the periodic deformation of the droplet by combining two aspects of its shape, $a_6$ and the droplet height~$h$.
For steady oscillations we display closed-loop trajectories for several frequencies~$f$ in Fig.~\ref{fig_windmill_2} and observe that their extent in the $h$-$a_6$ plane increases in all directions as $f$ decreases.
This is unlike the enclosed area for uniform wettability in Fig.~\ref{figure2}, which decreased for $f<10^{-3}\,\tau^{-1}$.

As mentioned before in Sect.~\ref{sec_pumping}, an increase in the enclosed area in configuration space means that the shape dynamics deviates more strongly from a reciprocal motion.
While for the oscillating uniform pattern we have the sperical cap model as a reference, which shows reciprocal dynamics and which the droplet should approach for sufficiently small $f$, such a reference is missing for the nonuniform patterns and the droplet dynamics need not be reciprocal in the limit of small $f$.

\section{Conclusions}
\label{sec_conclusions}
We have studied liquid droplets on substrates with oscillating wettability focussing on their shape and internal fluid flow.
When starting the wettability oscillations, the droplets go through a transient period, where the mean contact angle relaxes, and then settles into steady oscillations w.r.t.~shape and contact angle.
The amplitude of the contact-angle oscillations decreases with increasing frequency and, of course, increases with the amplitude of the wettability oscillations.
At small frequencies the amplitude and phase shift of the oscillations follow the linearized and parameter-free spherical-cap model, while they deviate from it for larger frequencies, where the free droplet surface deviates noticeably from a spherical cap.
As a result, amplitude and phase shift fall onto a master curve for small frequencies when we scale frequency with the decay time of the linearized spherical-cap model.

Upon further analysis of the droplet shape and the interior flow field, we have found that also for small frequencies the spherical cap model cannot account for the droplet's behavior completely.
The droplet shape deforms non-reciprocally which becomes evident when tracking its contact angle and height over time.
The non-reciprocal dynamics of the droplet shape gives rise to a time-dependent internal flow field which displaces point-like tracer particles over the course of a full period of oscillation rather than returning them to their original position.
Notably, the volume median of the displacement per period, the pumping speed, is constant w.r.t.~frequency in the investigated frequency range and only depends on material properties of the droplet and substrate.
Most importantly, the circulatory pumping of fluid inside the droplet is captured only by solving the full equations of Stokes flow, which we did using the boundary element method.

Repeating the same analysis for a droplet which is controlled by an oscillating nonuniform pattern of wettability with six-fold symmetry, we observed a very similar behavior.
The droplet settles into steady oscillations, the amplitude of which decreases when the oscillation frequency increases.
Furthermore, its shape deforms non-reciprocally, which becomes more pronounced  at smaller oscillation frequencies.

Because the circulation is stimulated from outside by oscillations in wettability, it provides a mechanism for controlling the transport and possibly mixing of solutes inside the droplet.
Internal transport through external stimuli has been used, for example, to precisely deposit a solute during the evaporation of droplets with light-responsive surfactants~\cite{varanakkottu_light_2016}.
Note, however, that depending on the contact angle evaporation contributes to and disturbs the flow field close to a moving contact line~\cite{gelderblom_stokes_2012}.
By designing specfic spatio-temporal wetttability patterns, such as moving steps in wettability~\cite{grawitter_steering_2021}, we envisage to similarly control the precise transport of solutes through substrates with switchable wettability.

\section*{Conflicts of Interest}
There are no conflicts to declare.

\appendix
\section{ESI}
\label{appendix_esi}
We provide
four example movies.
They show a droplet on a substrate, the wettability of which oscillates with frequency~$f=10^{-3}$. The first three movies show a 
droplet on a substrate with uniform wettability. Movie~M01 corresponds to an oscillation with
$\theta_\mathrm{eq}^\mathrm{min}=60^\circ$ and
$\theta_\mathrm{eq}^\mathrm{max}=120^\circ$,
movie~M02 corresponds to
$\theta_\mathrm{eq}^\mathrm{min}=45^\circ$ and
$\theta_\mathrm{eq}^\mathrm{max}=90^\circ$, and
movie~M03 corresponds to
$\theta_\mathrm{eq}^\mathrm{min}=30^\circ$ and
$\theta_\mathrm{eq}^\mathrm{max}=60^\circ$.
The final movie~M04 corresponds to the nonuniform wettability profile given by Eq.~(\ref{eq_nonuniform}) 
with $\theta_\mathrm{eq}^\mathrm{min}=60^\circ$ and $\theta_\mathrm{eq}^\mathrm{max}=120^\circ$.

\section*{Acknowledgements}
We acknowledge financial support from DFG (German Research Foundation) via Collaborative Research Center~910.


\printbibliography[heading=bibintoc]

@article{ichimura_light_2000,
    title   = {Light-Driven Motion of Liquids on a Photoresponsive Surface},
    author  = {Kunihiro Ichimura and Sangkeun Oh and Masaru Nakagawa},
    journal = {Science},
    volume  = {288},
    pages   = {1624},
    year    = {2000},
    doi     = {10.1126/science.288.5471.1624},
    eprinttype = {semanticscholar},
    eprint = {25350736}
}

@article{lahann_reversibly_2003,
    title   = {A Reversibly Switching Surface},
    author  = {Joerg Lahann and Samir Mitragotri and Thanhnga Tran and Hiroki Kaido and Jagannathan Sundaram and Insung S. Choi and Saskia Hoffer and Gabor A. Somorjai and Robert Langer},
    journal = {Science},
    volume  = {299},
    pages   = {371},
    year    = {2003},
    doi     = {10.1126/science.1078933}
}

@article{glasner_boundary_2005,
    title   = {A boundary integral formulation of quasi-steady fluid wetting},
    author  = {K. B. Glasner},
    journal = {J. Comp. Phys.},
    volume  = {207},
    pages   = {529},
    year    = {2005},
    doi     = {10.1016/j.jcp.2005.01.022}
}

@article{gelderblom_stokes_2012,
    title   = {Stokes flow near the contact line of an evaporating drop},
    author  = {Hanneke Gelderblom and Oscar Bloemen and Jacco H. Snoeijer},
    journal = {J. Fluid Mech.},
    volume  = {709},
    pages   = {69},
    year    = {2012},
    doi     = {10.1017/jfm.2012.321},
    eprinttype = {arxiv},
    eprint = {1111.6752}
}

@article{delorme_azobenzene_2005,
    title   = {Azobenzene-Containing Monolayer with Photoswitchable Wettability},
    author  = {N. Delorme and J.-F. Bardeau and A. Bulou and F. Poncin-Epaillard},
    journal = {Langmuir},
    volume  = {21},
    pages   = {12278},
    year    = {2005},
    doi     = {10.1021/la051517x}
}

@article{jiang_photoswitched_2005,
    title   = {Photo-switched wettability on an electrostatic self-assembly azobenzene monolayer},
    author  = {Wuhui Jiang and Guojie Wang and Yaning He and Xiaogong Wang and Yonglin An and Yanlin Songa and Lei Jiang},
    journal = {Chem. Commun.},
    pages   = {3550},
    year    = {2005},
    issue   = {28},
    doi     = {10.1039/b504479k}
}

@article{lim_photoreversibly_2006,
    title   = {Photoreversibly Switchable Superhydrophobic Surface with Erasable and Rewritable Pattern},
    author  = {Ho Sun Lim and Joong Tark Han and Donghoon Kwak and Meihua Jin and Kilwon Cho},
    journal = {J. Am. Chem. Soc.},
    volume  = {128},
    pages   = {14458},
    year    = {2006},
    doi     = {10.1021/ja0655901}
}

@article{baigl_photo_2012,
    title   = {Photo-actuation of liquids for light-driven microfluidics: state of the art and perspectives},
    author  = {Damien Baigl},
    journal = {Lab Chip},
    volume  = {12},
    pages   = {3637},
    year    = {2012},
    doi     = {10.1039/c2lc40596b}
}

@article{greenspan_motion_1978,
    title   = {On the motion of a small viscous droplet that wets a surface},
    author  = {H. P. Greenspan},
    journal = {J. Fluid. Mech.},
    volume  = {84},
    pages   = {125},
    year    = {1978},
    doi     = {10.1017/S0022112078000075}
}

@article{varanakkottu_light_2016,
    title   = {Light-Directed Particle Patterning by Evaporative Optical Marangoni Assembly},
    author  = {Subramanyan Namboodiri Varanakkottu and Manos Anyfantakis and Mathieu Morel and Sergii Rudiuk and Damien Baigl},
    journal = {Nano Lett.},
    volume  = {16},
    pages   = {644},
    year    = {2016},
    doi     = {10.1021/acs.nanolett.5b04377}
}

@article{kaspar_confinement_2016,
    title = {Confinement of water droplets on rectangular micro/nano-arrayed surfaces},
    author = {Kašpar, Ondřej and Zhang, Hailong and Tokárová, Viola and Boysen, Reinhard I. and Suñé, Gemma Rius and Borrise, Xavier and Perez-Murano, Francesco and Hearn, Milton T. W. and Nicolau, Dan V.},
    journal = {Lab Chip},
    volume = {16},
    pages = {2487},
    year = {2016},
    doi = {10.1039/c6lc00622a}
}

@article{bunker_reversible_2008,
    title = {Reversible switching of interfacial interactions},
    author = {Bruce C. Bunker},
    journal = {Mater. Sci. Eng. R},
    volume = {62},
    pages = {157},
    year = {2008},
    doi = {10.1016/j.mser.2008.06.001}
}

@book{kim_microhydrodynamics_2005,
    title = {Microhydrodynamics},
    author = {Sangtae Kim and Sepp J. Karrila},
    year = {2005},
    isbn = {0486442195},
    publisher = {Dover Publications},
    address = {Mineola/NY}
}

@article{bonn_wetting_2009,
    title = {Wetting and spreading},
    author = {Daniel Bonn and Jens Eggers and Joseph Indekeu and Jacques Meunier and Etienne Rolley},
    journal = {Rev. Mod. Phys.},
    volume = {81},
    pages = {739},
    year = {2009},
    doi = {10.1103/RevModPhys.81.739}
}

@article{huh_hydrodynamic_1971,
    title = {Hydrodynamic Model of Steady Movement of a Solid / Liquid / Fluid Contact Line},
    author = {Chun Huh and L. E. Scriven},
    journal = {J. Colloid Interface Sci.},
    volume = {35},
    pages = {85},
    year = {1971},
    doi = {10.1016/0021-9797(71)90188-3}
}

@article{brochard_motions_1989,
    title = {Motions of Droplets on Solid Surfaces Induced by Chemical or Thermal Gradients},
    author = {F. Brochard},
    journal = {Langmuir},
    volume = {5},
    pages = {432},
    year = {1989},
    doi = {10.1021/la00086a025}
}

@article{oron_long_1997,
    title = {Long-scale evolution of thin liquid films},
    author = {Alexander Oron and Stephen H. Davis and S. George Bankoff},
    journal = {Rev. Mod. Phys.},
    volume = {69},
    pages = {931},
    year = {1997},
    doi = {10.1103/RevModPhys.69.931}
}

@article{thiele_dynamical_2004,
    title = {Dynamical Model for Chemically Driven Running Droplets},
    author = {Uwe Thiele and Karin John and Markus Bär},
    journal = {Phys. Rev. Lett.},
    volume = {93},
    pages = {027802},
    year = {2004},
    doi = {10.1103/PhysRevLett.93.027802},
    eprinttype = {arxiv},
    eprint = {cond-mat/0403043}
}

@book{pozrikidis_boundary_1992,
    title = {Boundary integral and singularity methods for linearized viscous flow},
    author = {C. Pozrikidis},
    publisher = {Cambridge University Press},
    address = {Cambridge},
    isbn = {9780521406932},
    year = {1992}
}

@article{eral_contact_2013,
    title = {Contact angle hysteresis: a review of fundamentals and applications},
    author = {H. B. Eral and D. J. C. M. {'t Mannetje} and J. M. Oh},
    journal = {Colloid Polym. Sci.},
    volume = {291},
    pages = {247},
    year = {2013},
    doi = {10.1007/s00396-012-2796-6},
    url = {https://dspace.mit.edu/bitstream/1721.1/86387/2/Eral_Contact%20angle.pdf},
    url+an = {=openaccess}
}

@article{voinov_hydrodynamics_1976,
    title = {Hydrodynamics of Wetting},
    author = {O. V. Voinov},
    journal = {Fluid Dyn.},
    volume = {11},
    pages = {714},
    year = {1976},
    doi = {10.1007/BF01012963}
}

@article{cox_dynamics_1986,
    title = {The dynamics of the spreading of liquids on a solid surface},
    author = {R. G. Cox},
    journal = {J. Fluid Mech.},
    volume = {168},
    pages = {169},
    year = {1986},
    doi = {10.1017/S0022112086000332}
}

@article{snoeijer_moving_2013,
    title = {Moving Contact Lines: Scales, Regimes, and Dynamical ransitions},
    author = {Snoeijer, Jacco H. and Andreotti, Bruno},
    journal = {Annu. Rev. Fluid Mech.},
    volume = {45},
    pages = {269},
    year = {2013},
    doi = {10.1146/annurev-fluid-011212-140734},
    eprinttype = {semanticscholar},
    eprint = {121130784}
}

@article{allen_orbital_1992,
    title = {Orbital angular momentum of light and the transformation of Laguerre-Gaussian laser modes},
    author = {L. Allen and M. W. Beijersbergen and R. J. C. Spreeuw and J. P. Woerdman},
    journal = {Phys. Rev. A},
    volume = {45},
    pages = {8185},
    year = {1992},
    doi = {10.1103/PhysRevA.45.8185}
}

@article{pityuk_boundary_2018,
    title = {Boundary Element Modeling of Dynamics of a Bubble in Contact with a Solid Surface at Low Reynolds Numbers},
    author = {Yu. A. Pityuk and O. A. Abramova and N. A. Gumerov and I. Sh. Akhatov},
    journal = {Math. Models Comput. Simul.},
    volume = {10},
    pages = {209},
    year = {2018},
    doi = {10.1134/S2070048218020102}
}

@article{tsitouras_runge_2011,
    title = {Runge–Kutta pairs of order 5(4) satisfying only the first column simplifying assumption},
    author = {Tsitouras, {Ch.}},
    journal = {Comput. Math. Appl.},
    volume = {62},
    pages = {770},
    year = {2011},
    doi = {10.1016/j.camwa.2011.06.002},
    doi+an = {=openaccess}
}

@article{moffatt_viscous_1964,
    title = {Viscous and resistive eddies near a sharp corner},
    author = {H. K. Moffatt},
    journal = {J. Fluid Mech.},
    volume = {18},
    pages = {1-18},
    year = {1964},
    doi = {10.1017/S0022112064000015}
}

@article{mcgraw_slip_2016,
    title = {Slip-mediated dewetting of polymer microdroplets},
    author = {Joshua D. McGraw and Tak Shing Chan and Simon Maurer and Thomas Salez and Michael Benzaquen and Elie Raphaël and Martin Brinkmann and Karin Jacobs},
    journal = {Proc. Natl. Acad. Sci. U.S.A.},
    volume = {113},
    pages = {1168},
    year = {2016},
    doi = {10.1073/pnas.1513565113},
    doi+an  = {=openaccess}
}

@article{bolanos_derivation_2017,
    title = {Derivation of the Navier slip and slip length for viscous flows over a rough boundary},
    author = {Silvia Jiménez Bolaños and Bogdan Vernescu},
    journal = {Phys. Fluids},
    volume = {29},
    pages = {057103},
    year = {2017},
    doi = {10.1063/1.4982899}
}

@article{deruijter_contact_1997,
    title = {Contact Angle Relaxation during the Spreading of Partially Wetting Drops},
    author = {de Ruijter, M. J. and De Coninck, J. and Blake, T. D. and Clarke, A. and Rankin, A.},
    journal = {Langmuir},
    volume = {13},
    pages = {7293},
    year = {1997},
    doi = {10.1021/la970825v}
}

@book{katsikadelis_boundary_2016,
    title = {The Boundary Element Method for Engineers and Scientists},
    author = {John T. Katsikadelis},
    publisher = {Elsevier},
    year = {2016},
    isbn = {978-0-12-804493-3},
    edition = {2}
}

@article{pirani_light_2016,
    title = {Light-Driven Reversible Shaping of Individual Azopolymeric Micro-Pillars},
    author = {Federica Pirani and Angelo Angelini and Francesca Frascella and Riccardo Rizzo and Serena Ricciardi and Emiliano Descrovi},
    journal = {Sci. Rep.},
    volume = {6},
    pages = {31702},
    year = {2016},
    doi = {10.1038/srep31702},
    doi+an  = {=openaccess}
}

@article{savva_droplet_2019,
    title = {Droplet dynamics on chemically heterogeneous substrates},
    author = {Nikos Savva and Danny Groves and Serafim Kalliadasis},
    journal = {J. Fluid Mech.},
    volume = {859},
    pages = {321},
    year = {2019},
    doi = {10.1017/jfm.2018.758},
    url = {http://orca.cf.ac.uk/118322/1/Savva_3DDrops.pdf},
    url+an = {=openaccess}
}

@article{chan_morphological_2017,
    title = {Morphological evolution of microscopic dewetting droplets with slip},
    author = {Chan, Tak Shing and McGraw, Joshua D. and Salez, Thomas and Seemann, Ralf and Brinkmann, Martin},
    journal = {J. Fluid Mech.},
    volume = {828},
    pages = {271},
    year = {2017},
    doi = {10.1017/jfm.2017.515},
    url = {https://eprints.lib.hokudai.ac.jp/dspace/bitstream/2115/68436/3/1612.01346_2.pdf},
    url+an = {=openaccess}
}

@article{squires_microfluidics_2005,
    title = {Microfluidics:~Fluid physics at the nanoliter scale},
    author = {Todd M. Squires and Stephen R. Quake},
    journal = {Rev. Mod. Phys.},
    volume = {77},
    pages = {977},
    year = {2005},
    doi = {10.1103/RevModPhys.77.977},
    url = {https://authors.library.caltech.edu/1310/1/SQUrmp05.pdf},
    url+an = {=openaccess}
}

@article{grawitter_steering_2021,
    title = {Steering droplets on substrates using moving steps in wettability},
    author = {Josua Grawitter and Holger Stark},
    journal = {Soft Matter},
    volume = {17},
    pages = {2454},
    year = {2021},
    doi = {10.1039/D0SM02082F},
    doi+an  = {=openaccess}
}

@article{purcell_life_1977,
    title = {Life at low Reynolds number},
    author = {E. M. Purcell},
    journal = {Am. J. Phys.},
    volume = {45},
    pages = {3},
    year = {1977},
    doi = {10.1119/1.10903}
}

@article{gauger_numerical_2006,
    title = {Numerical study of an artificial microswimmer},
    author = {E. Gauger and Holger Stark},
    journal = {Phys. Rev. E},
    year = {2006},
    volume = {74},
    pages = {021907},
    doi = {10.1103/PhysRevE.74.021907},
    eprinttype = {arxiv},
    eprint = {0805.3137}
}

@article{seki_wide_2018,
    title = {A Wide Array of Photoinduced Motions in Molecular and Macromolecular Assemblies at Interfaces},
    author = {Takahiro Seki},
    journal = {Bull. Chem. Soc. Jpn.},
    volume = {91},
    number = {7},
    pages = {1026-1057},
    year = {2018},
    doi+an = {=openaccess},
    doi = {10.1246/bcsj.20180076}
}

@article{malinowski_advances_2020,
    title = {Advances towards programmable droplet transport on solid surfaces and its applications},
    author = {Robert Malinowski and Ivan P. Parkin and Giorgio Volpe},
    journal = {Chem. Soc. Rev.},
    volume = {49},
    pages = {7879},
    year = {2020},
    doi+an = {=openaccess},
    doi = {10.1039/D0CS00268B}
}

@article{young_essay_1805,
    author = {Young, Thomas },
    title = {An essay on the cohesion of fluids},
    journal = {Philos. Trans. R. Soc.},
    volume = {95},
    pages = {65-87},
    year = {1805},
    doi = {10.1098/rstl.1805.0005}
}

@article{lim_uvdriven_2007,
    title = {UV-Driven Reversible Switching of a Roselike Vanadium Oxide Film between Superhydrophobicity and Superhydrophilicity},
    author = {Ho Sun Lim and Donghoon Kwak and Dong Yun Lee and Seung Goo Lee and Kilwon Cho},
    journal = {J. Am. Chem. Soc.},
    volume = {129},
    pages = {4128},
    year = {2007},
    doi = {10.1021/ja0692579}
}

@article{oscurato_light_2017,
    author = {Oscurato, Stefano Luigi and Borbone, Fabio and Maddalena, Pasqualino and Ambrosio, Antonio},
    title = {Light-Driven Wettability Tailoring of Azopolymer Surfaces with Reconfigured Three-Dimensional Posts},
    journal = {ACS Appl. Mater. Interfaces},
    volume = {9},
    number = {35},
    pages = {30133-30142},
    year = {2017},
    doi = {10.1021/acsami.7b08025}
}

@article{teletzke_wetting_1988,
    title = {Wetting hydrodynamics},
    author = {Gary F. Teletzke and H. Ted Davis and L.E. Scriven},
    journal = {Rev. Phys. appl.},
    volume = {23},
    pages = {989-1007},
    year = {1988},
    doi = {10.1051/rphysap:01988002306098900},
    eprinttype = {hal},
    eprint = {jpa-00245927}
}

@article{popescu_precursor_2012,
    title = {Precursor films in wetting phenomena},
    author = {M. N. Popescu and G. Oshanin and S. Dietrich and A.-M. Cazabat},
    journal = {J. Phys.: Condens. Matter},
    volume = {24},
    pages = {243102},
    year = {2012},
    doi = {10.1088/0953-8984/24/24/243102},
    eprinttype = {arxiv},
    eprint = {1205.1541}
}

@article{thiele_recent_2018,
    title = {Recent advances in and future challenges for mesoscopic hydrodynamicmodelling of complex wetting},
    author = {Uwe Thiele},
    journal = {Colloids Surf. A},
    volume = {553},
    pages = {487},
    year = {2018},
    doi = {10.1016/j.colsurfa.2018.05.049},
    eprinttype = {arxiv},
    eprint = {1803.05388}
}

@article{sudo_dynamic_2010,
  title={The Dynamic Behavior of Liquid Droplets on Vibrating Plate},
  author={Seiichi Sudo and Ayaka Goto and Hiroki Kuwano and Yuichiro Hamate and Tetsuya Yano and Kyohei Hoshika},
  journal={J. Jpn. Soc. Exp. Mech.},
  volume={10},
  pages={s38-s45},
  year={2010},
  doi={10.11395/jjsem.10.s38}
}

@article{singla_dynamics_2019,
    author = {Singla,Tanu  and Verma,Dinesh Kumar  and Tovar,Josué Flores  and Figueroa,A.  and Vázquez,Federico  and Yousif,Farook Bashir  and Rivera,M. },
    title = {Dynamics of a vertically vibrating mercury drop},
    journal = {AIP Adv.},
    volume = {9},
    number = {4},
    pages = {045204},
    year = {2019},
    doi = {10.1063/1.5088043},
    doi+an = {=openaccess}
}

@article{noblin_vibrations_2009,
    title = {Vibrations of sessile drops},
    author = {Noblin, X. and Buguin, A. and Brochard-Wyart, F.},
    journal = {Eur. Phys. J. Spec. Top.},
    volume = {166},
    pages = {7-10},
    year = {2009},
    doi = {10.1140/epjst/e2009-00869-y}
}

@article{dai_directional_2021,
    title = {Directional interfacial motion of liquids: Fundamentals, evaluations, and manipulation strategies},
    journal = {Tribol. Int.},
    volume = {154},
    pages = {106749},
    year = {2021},
    issn = {0301-679X},
    doi = {10.1016/j.triboint.2020.106749},
    author = {Qingwen Dai and Wei Huang and Xiaolei Wang and M.M. Khonsari}
}

@article{dietrich_wetting_2005,
    title = {Wetting on structured substrates},
    author = {Siegfried Dietrich and M. N. Popescu and M. Rauscher},
    journal = {J. Phys.: Condens. Matter},
    volume = {17},
    pages = {S577},
    year = {2005},
    doi = {10.1088/0953-8984/17/9/017}
}

@article{moosavi_size_2008,
    author = {Moosavi,A.  and Rauscher,M.  and Dietrich,S. },
    title = {Size dependent motion of nanodroplets on chemical steps},
    journal = {J. Chem. Phys.},
    volume = {129},
    number = {4},
    pages = {044706},
    year = {2008},
    doi = {10.1063/1.2955860},
    eprinttype = {arxiv},
    eprint = {0805.0506}
}

@article{degennes_wetting_1985,
    author = {Pierre-Gilles de Gennes},
    title = {Wetting: statics and dynamics},
    journal = {Rev. Mod. Phys.},
    volume = {57},
    pages = {827},
    year = {1985},
    doi = {10.1103/RevModPhys.57.827}
}

@article{samiei_review_2016,
    author ="Samiei, Ehsan and Tabrizian, Maryam and Hoorfar, Mina",
    title  ="A review of digital microfluidics as portable platforms for lab-on a-chip applications",
    journal  ="Lab Chip",
    year  ="2016",
    volume  ="16",
    issue  ="13",
    pages  ="2376-2396",
    publisher  ="The Royal Society of Chemistry",
    doi  ="10.1039/C6LC00387G",
    url = "https://escholarship.mcgill.ca/concern/articles/vt150p551",
    url+an = "=openaccess"
}

@article{mampallil_review_2018,
    title = {A review on suppression and utilization of the coffee-ring effect},
    journal = {Advances in Colloid and Interface Science},
    volume = {252},
    pages = {38-54},
    year = {2018},
    issn = {0001-8686},
    doi = {10.1016/j.cis.2017.12.008},
    author = {Dileep Mampallil and Huseyin Burak Eral},
    url = {http://resolver.tudelft.nl/uuid:b12d4d79-67af-4177-b03c-88bfaa8b622d},
    url+an = {=openaccess}
}

@inbook{engelnkemper_continuation_2019,
author="Engelnkemper, S.
and Gurevich, S. V.
and Uecker, H.
and Wetzel, D.
and Thiele, U.",
editor="Gelfgat, Alexander",
title="Continuation for Thin Film Hydrodynamics and Related Scalar Problems",
bookTitle="Computational Modelling of Bifurcations and Instabilities in Fluid Dynamics",
year="2019",
publisher="Springer International Publishing",
address="Cham",
pages="459--501",
isbn="978-3-319-91494-7",
doi="10.1007/978-3-319-91494-7_13",
eprinttype="arxiv",
eprint="1808.02321"
}
\end{document}